\documentclass[aps,prl,reprint,superscriptaddress,floatfix]{revtex4-2}

\usepackage{graphicx,amsmath}
\usepackage{hyperref}
\usepackage{siunitx}
\hypersetup{colorlinks=true, citecolor=blue, urlcolor=blue, linkcolor=blue}
\usepackage{xcolor}

\usepackage{pdfpages} 
\usepackage{pgffor} 

\makeatletter
\AtBeginDocument{\let\LS@rot\@undefined}
\makeatother

\def\supplementfilename{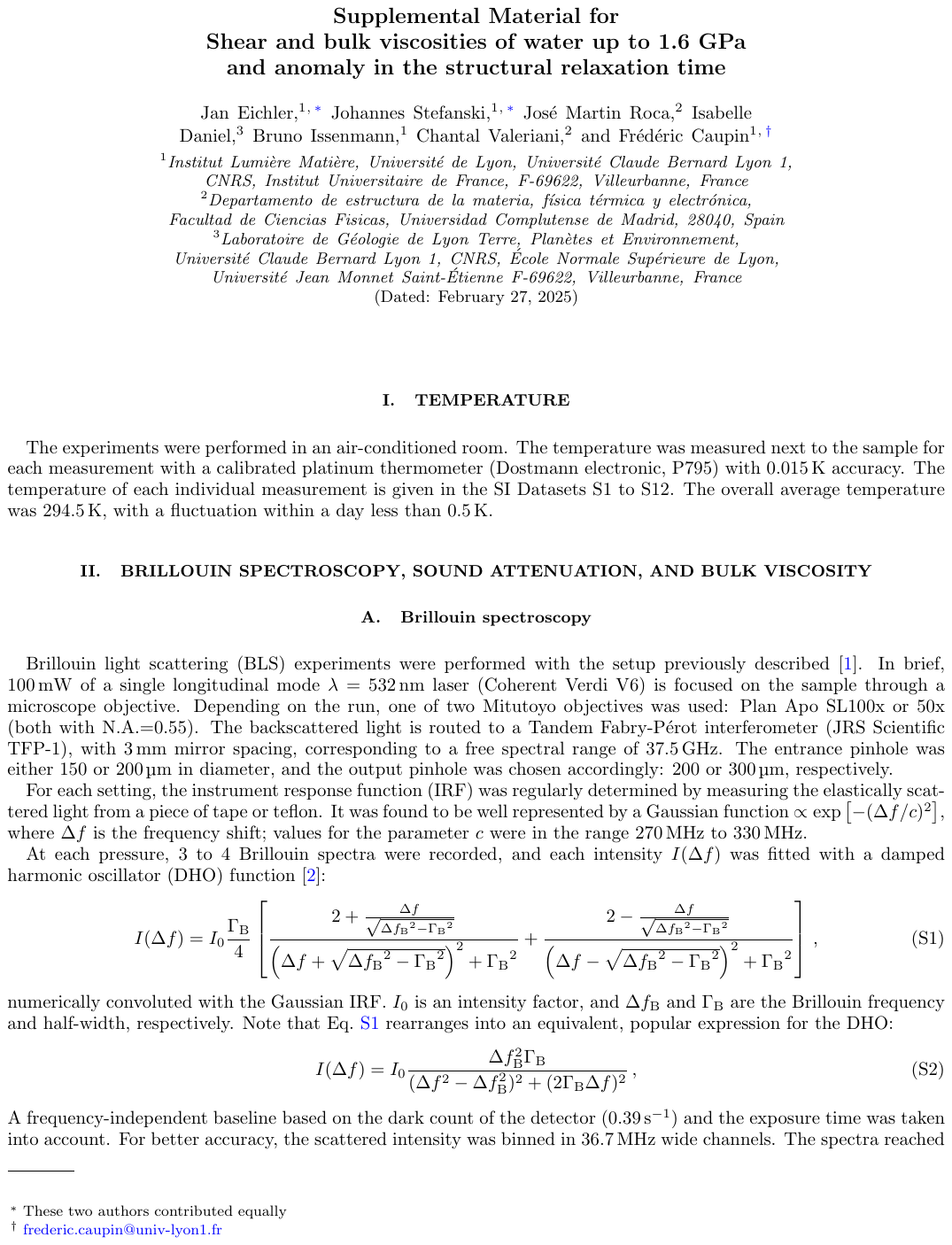}

\pdfximage{\supplementfilename}
\def\numbersupplementpages{\the\pdflastximagepages}

\newif\ifarXiv
\arXivtrue 

\begin{document}


\title{Shear and bulk viscosities of water up to 1.6 GPa\\
and anomaly in the structural relaxation time}


\author{Jan Eichler}\thanks{These two authors contributed equally}
\author{Johannes Stefanski}\thanks{These two authors contributed equally}
\affiliation{Institut Lumi\`ere Mati\`ere, Universit\'e de Lyon, Universit\'e Claude Bernard Lyon 1, CNRS, Institut Universitaire de France, F-69622, Villeurbanne, France}
\author{Jos\'{e} Martin Roca}
\affiliation{Departamento de estructura de la materia, f\'{\i}sica t\'ermica y electr\'onica, Facultad de Ciencias Fisicas, Universidad Complutense de Madrid, 28040, Spain}
\author{Isabelle Daniel}
\affiliation{Laboratoire de G\'eologie de Lyon Terre, Plan\`etes et Environnement, Universit\'e Claude Bernard Lyon 1, CNRS, \'Ecole Normale Sup\'erieure de Lyon, Universit\'e Jean Monnet Saint-\'Etienne F-69622, Villeurbanne, France}
\author{Bruno Issenmann}
\affiliation{Institut Lumi\`ere Mati\`ere, Universit\'e de Lyon, Universit\'e Claude Bernard Lyon 1, CNRS, Institut Universitaire de France, F-69622, Villeurbanne, France}
\author{Chantal Valeriani}
\affiliation{Departamento de estructura de la materia, f\'{\i}sica t\'ermica y electr\'onica, Facultad de Ciencias Fisicas, Universidad Complutense de Madrid, 28040, Spain}
\author{Fr\'{e}d\'{e}ric Caupin}
\email[]{frederic.caupin@univ-lyon1.fr}
\affiliation{Institut Lumi\`ere Mati\`ere, Universit\'e de Lyon, Universit\'e Claude Bernard Lyon 1, CNRS, Institut Universitaire de France, F-69622, Villeurbanne, France}


\date{February 27, 2025}

\begin{abstract}
Deep in the Earth's crust, pressure exceeds one thousand times the atmospheric pressure. Water still flows under these conditions, but experiences dramatic changes in structure and fluidity. Using combined dynamic and inelastic light scattering techniques, we simultaneously measure the shear and bulk viscosities of water as a function of pressure. The former increases faster than the latter, so that their ratio shows a two-fold decrease from 0 to 1.6 GPa; we confirm this trend with simulations. We analyze our results in terms of the structural relaxation time $\tau$. Contrary to other liquids, pressure initially accelerates relaxation in water. Our measurements reveal that $\tau$ reaches a minimum close to \qty{1}{ps} around \qty{0.5}{GPa}. We interpret $\tau$ as a the equilibration time of hydrogen bonds, and propose that the minimum in $\tau$ arises from a structural anomaly which allows fastest interconversion between local structures in water, and generates a cascade of thermodynamic and dynamic anomalies.
\end{abstract}


\maketitle



Water is a very anomalous liquid~\cite{gallo_water_2016}. Its famous oddities, e.g. a negative isobaric expansion coefficient~\cite{mishima_volume_2010}, or an initial decrease of viscosity with increasing pressure~\cite{singh_pressure_2017,mussa_viscosity_2023}, grow upon cooling, but disappear at sufficiently high pressures, above which water adopts a thermodynamic and dynamic behavior similar to simple liquids. A key to water's anomalies is given by its very open structure near ambient conditions. Hydrogen bonds between molecules favor a tetrahedral arrangement of the first four nearest neighbors around a given molecule, and push the fifth nearest neighbor to a rather large distance. Applying pressure distorts the hydrogen bond network, moving the fifth neighbor to shorter distances~\cite{soper_structures_2000}. This structural change occurs progressively over a pressure range of several hundreds of megapascals, and may affect various properties differently. The present work focuses on the behavior of viscosity at ambient temperature. Viscosity is a macroscopic quantity, but reflects microscopic structural properties: for instance, in recent optical Kerr effect experiments~\cite{kacenauskaite_fast_2024}, the viscosity of HCl solutions in water was found to linearly correlate with the slowest component of dynamic relaxation, suggesting that viscosity in these solutions is determined by the structural dynamics of the hydronium/Cl\textsuperscript{-}/water configurations. Here, we study pure water from \num{0} to \qty{1.6}{GPa}. This spans a density range from \num{1000} to \qty{1300}{kg.m^{-3}}, where the liquid becomes metastable with respect to ice VI. The structural relaxation time $\tau$ in liquids usually increases with pressure, but water shows the opposite behavior~\cite{bencivenga_temperature_2009}. Our goal is to use viscosity to find if, and at which pressure, this anomaly disappears, which would signal a pivotal structural change.

In addition to its relation to the liquid structure, the viscosity at high pressure is involved in technology and natural phenomena. For instance, water-flooding of oil reservoirs is a secondary oil recovery method which consists in injecting water to push oil out. The increased depth of new wells involves pressures exceeding \qty{240}{MPa}, and knowledge of fluid properties at these conditions is required for process optimization and safety~\cite{mallepally_fluid_2018}. Life, for which liquid water is a prerequisite, is also found at high pressure. Earth has been inhabited for more than 3 billion years, with most of its biomass located in the subsurface~\cite{bar-on_biomass_2018} and cells potentially present at $\simeq$\qty{25}{\km} depth or pressure of \qty{0.7}{GPa}~\cite{magnabosco_biomass_2018}. Small increases in the viscosity of water counter-intuitively increase cell motility to some extent~\cite{nsamela_effect_2021}, which may have consequences on the habitability of subsurface environments with low geothermal gradients such as beneath cratons, mud volcanoes, where percolation in porous media is the dominant transfer mode. In the solar system, a series of moons and asteroids show strong evidence of liquid water ocean beneath their icy shells, such as Ganymede, in which melting may occur at around 1.6 GPa at the interface between the hydrosphere and the silicate mantle. Beyond the solar system, some smaller or larger exomoons with subsurface ocean are expected to exist as well~\cite{heller_water_2015}. Assessing the transfer of chemical energy and nutrient sources requires convection models. As of today, the most sophisticated ones include a fixed value of the shear viscosity of water~\cite{kalousova_twophase_2018}. Any increase in viscosity will slow down the percolation of water in the icy subsurface. However, any effect on the convection would require changes in the viscosity of a few orders of magnitude.

Yet, the viscosity of water in the gigapascal range has been scarcely measured, and large discrepancies exist between published data~\cite{abramson_viscosity_2007,bowman_optical_2013,frost_isotope_2020}. Here, to measure shear viscosity at high pressure, we adapted to a diamond anvil cell (DAC)~\cite{SM} differential dynamic microscopy (DDM)~\cite{cerbino_differential_2008}, a technique successfully used before at low pressure in supercooled water~\cite{dehaoui_viscosity_2015,ragueneau_shear_2022,mussa_viscosity_2023}. DDM relies on the Brownian motion of freely suspended sub-micrometer spheres. In addition, we carried simultaneous Brillouin spectroscopy measurements\cite{SM}. When a laser illuminates the sample, light is scattered inelastically by density fluctuations. The linewidth of the scattered light provides the sound attenuation. Sound attenuation cannot be accounted for solely by shear viscosity. In Newtonian fluids, a second parameter, called bulk or volume viscosity, is required, as recognized in 1845 by Stokes~\cite{stokes_theories_1845}. Our joint DDM and Brillouin measurements enable access to the separate values of shear and bulk viscosities. We complement our measurements with molecular dynamics simulations of the same quantities. We analyze our results with a visco-elastic model~\cite{litovitz_structural_1965,anisimov_mesoscopic_2024} which yields the structural relaxation time of water.


\begin{figure}[!t]
\centering
\includegraphics[width=.95\linewidth]{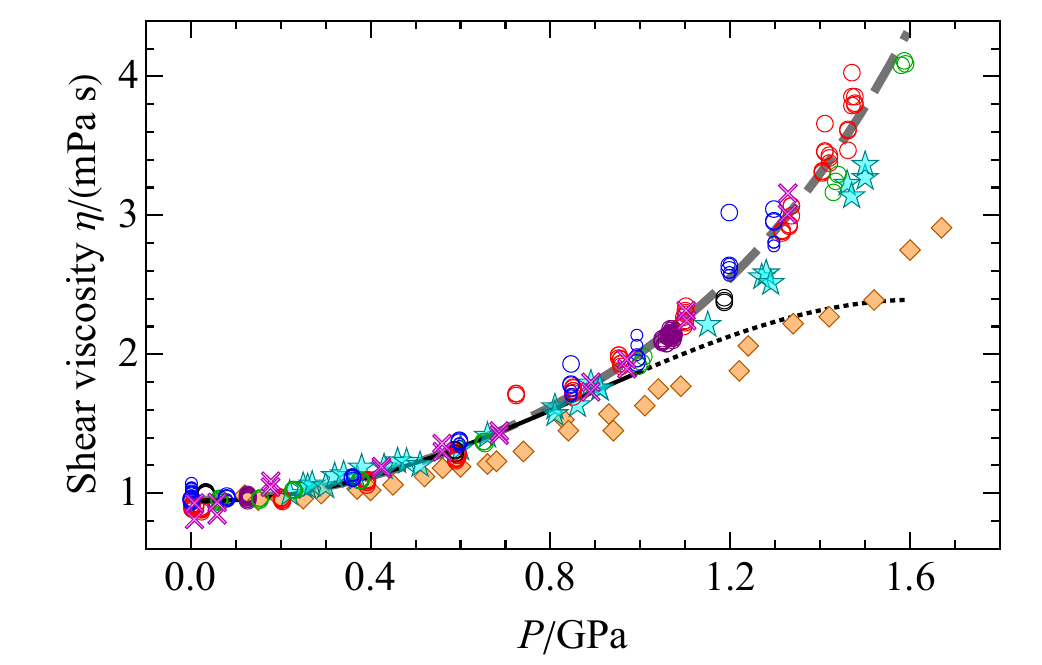}
\caption{\textbf{Shear viscosity of water as a function of pressure}. This work: open circles for runs A (blue), B (red), C (black), D (purple), and E (green). The large and small symbols stand for data obtained with 100x and 50x objectives, respectively. Also shown are the data obtained with the rolling ball method~\cite{abramson_viscosity_2007} (cyan stars), optical tweezers~\cite{bowman_optical_2013} (magenta crosses), and a previous DDM attempt~\cite{frost_isotope_2020} (orange diamonds). The solid and dotted black curves represent the IAPWS formulation for viscosity~\cite{huber_new_2009} at \qty{295.16}{K} and its extrapolation above \qty{1}{\GPa}, respectively. The thick, dashed, gray curve is a $4^\mathrm{th}$ order polynomial fit to all our data, whose parameters are given in Supplementary Material (SM)~\cite{SM}.
\label{fig:eta}
}
\end{figure}
\textit{Shear viscosity---}Figure~\ref{fig:eta} shows our experimental results at \qty{295}{K} for shear viscosity $\eta$ for a total of 5 runs; for two of them, measurements were performed with two objectives with different magnification. Overall, there is excellent agreement between all our measurements. The standard deviation around a smooth curve is 4.3\%, only slightly higher than the typical \numrange{2}{3}\% random error for DDM measurements~\cite{dehaoui_viscosity_2015}. The pressure dependence is initially flat. Actually, there is a shallow minimum at ambient temperature, which gets more pronounced upon cooling~\cite{singh_pressure_2017,mussa_viscosity_2023}, but the present data scatter does not allow resolving it. Then, $\eta$ increases more and more rapidly with pressure. Our data are consistent with the IAPWS formulation for viscosity~\cite{huber_new_2009}, up to its stated limit of validity of \qty{1}{GPa}, but departs upward from its extrapolation to higher pressures. Only a few other works were carried under high pressure. Abramson's data~\cite{abramson_viscosity_2007} fall systematically below ours at higher pressure. We do not have a clear explanation for this moderate discrepancy (see SM for a discussion possible biases in the rolling ball method). Our data are fully consistent with Ref.~\cite{bowman_optical_2013}, obtained with an independent technique. Previous DDM data~\cite{frost_isotope_2020} are noticeably lower than all other works. We suspect this may be caused by rapid sedimentation of the silica spheres used in that study. Appendix A discusses further the comparison with the literature.

\begin{figure}[!b]
\centering
\includegraphics[width=.95\linewidth]{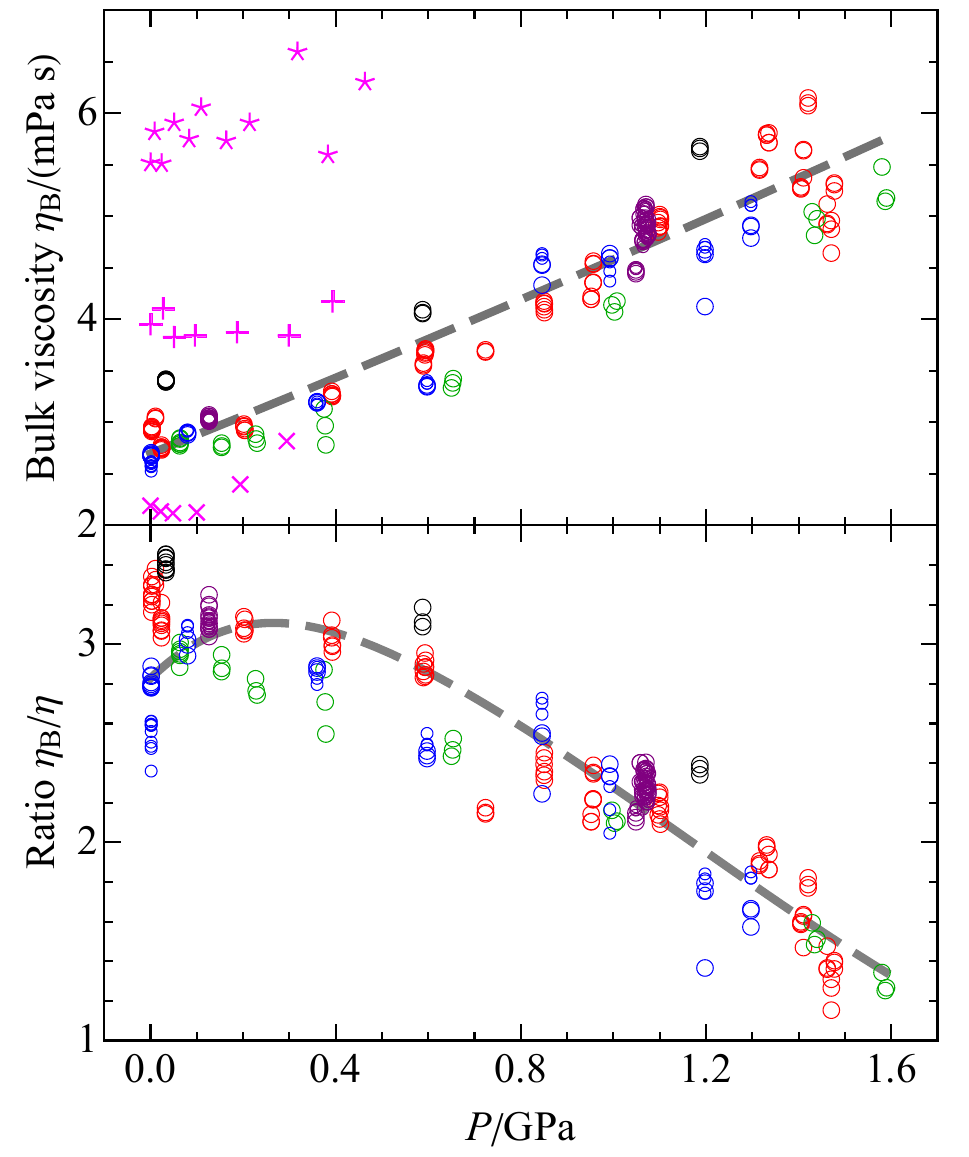}
\caption{\textbf{Bulk viscosity of water under pressure.} Top panel: bulk viscosity as a function of pressure. Our results are shown with open circles for runs A (blue), B (red), C (black), D (purple), and E (green). The thick, dashed, gray curve is a second-order polynomial fit to all our data, whose parameters are given in SM. Ultrasonic data~\cite{hawley_ultrasonicabsorption_1970} at \num{273}, \num{283.3}, and \qty{303}{K} are shown with magenta asterisks, pluses, and crosses, respectively. Bottom panel: ratio between bulk and shear viscosity as a function of pressure. The thick, dashed, gray curve shows the ratio of the respective fits to all our data (top panels of Figs.~\ref{fig:bulk} and \ref{fig:eta}).
\label{fig:bulk}
}
\end{figure}

\textit{Bulk viscosity---}
The sound attenuation coefficient $\alpha/f^2$ measured with Brillouin is consistent with ultrasonic experiments~\cite{litovitz_effect_1955,hawley_ultrasonicabsorption_1970} and extends to higher pressure (see Appendix B). The combined knowledge of $\eta$ and $\alpha/f^2$ gives access to bulk viscosity $\eta_\mathrm{b}$ using:
\begin{equation}
\frac{\alpha}{f^2} = \frac{2 \pi^2}{\rho w^3} \left( \frac{4}{3} \eta + \eta_\mathrm{b} \right) \, ,
\label{eq:attenuation}
\end{equation}
where $\rho$ is the density, $w$ is the sound velocity, and a term involving thermal conductivity has been neglected (see justification in SM). Results for $\eta_\mathrm{b}$ from five runs are displayed in Fig.~\ref{fig:bulk}. They are fairly consistent with each other, showing a regular increase with pressure, with a standard deviation of 7.5 \% around a parabolic fit. The data from Ref~\cite{hawley_ultrasonicabsorption_1970} at neighboring temperatures bracket our data. They reveal a shallow minimum in $\eta_\mathrm{b}$, which the scatter of our data does not allow resolving. Nevertheless, they demonstrate an interesting behavior of the ratio $\eta_\mathrm{b}/\eta$ (Fig.~\ref{fig:bulk}, bottom): first roughly constant up to \qty{0.6}{\GPa}, it starts decreasing at higher pressure, reaching at \qty{1.6}{\GPa} only one half of the ambient pressure value. The pressure effect is rather large. For comparison, $\eta_\mathrm{b}/\eta$ increases only from \num{2.7} to \num{3.1} when cooling from \num{323.15} to \qty{280.15}{\K}~\cite{holmes_temperature_2011}, and decreases again to around \num{2.5} at \qty{253.15}{K}~\cite{conde_analysis_1982}. We have performed simulations of shear and bulk viscosity for the TIP4P/2005 model of water~\cite{abascal_general_2005}: they confirm the observed experimental trend (see Appendix B for details).

\textit{Visco-elastic interpretation and relaxation time---}Sound attenuation in liquids is intimately linked to microscopic structural relaxation, characterized by a structural relaxation time $\tau$~\cite{litovitz_structural_1965,anisimov_mesoscopic_2024}. At low enough frequencies ($\ll 1/\tau$), sound waves propagate at a constant velocity $w_0 $. When the sound wave frequency approaches $1/\tau$, microscopic relaxation cannot occur sufficiently fast during the sound period. This induces a time lag in the system's response, which causes a peak in dissipation. In turn, the dissipation peak induces a sharp increase in the sound velocity. Eventually, at very high frequencies, the system behaves like an elastic solid, with low dissipation and sound propagation at a limiting frequency $w_\infty$. Inelastic x-ray scattering probes the high frequency limit and provides $w_\infty$, for which values are available in water near ambient temperature from 0 to \qty{1.35}{GPa}~\cite{krisch_pressure_2002} (SM). This is the basis of the visco-elastic relaxation model of liquids. It introduces adiabatic elastic moduli for the liquid: $K_0=\rho {w_0}^2$ and $K_\infty$ are the bulk moduli at zero and infinite frequency, respectively, and $G_\infty$ the shear modulus at infinite frequency. Assuming a single relaxation process (with the same characteristic time $\tau$ for shear and bulk relaxation at constant pressure, an assumption supported by experiments~\cite{slie_ultrasonic_1966,darrigo_sound_1981} and simulations~\cite{bertolini_stress_1995}), it yields the following relations at low frequencies~\cite{litovitz_structural_1965,magazu_relaxation_1989,anisimov_mesoscopic_2024} :
\begin{figure}[!t]
\centering
\includegraphics[width=.95\linewidth]{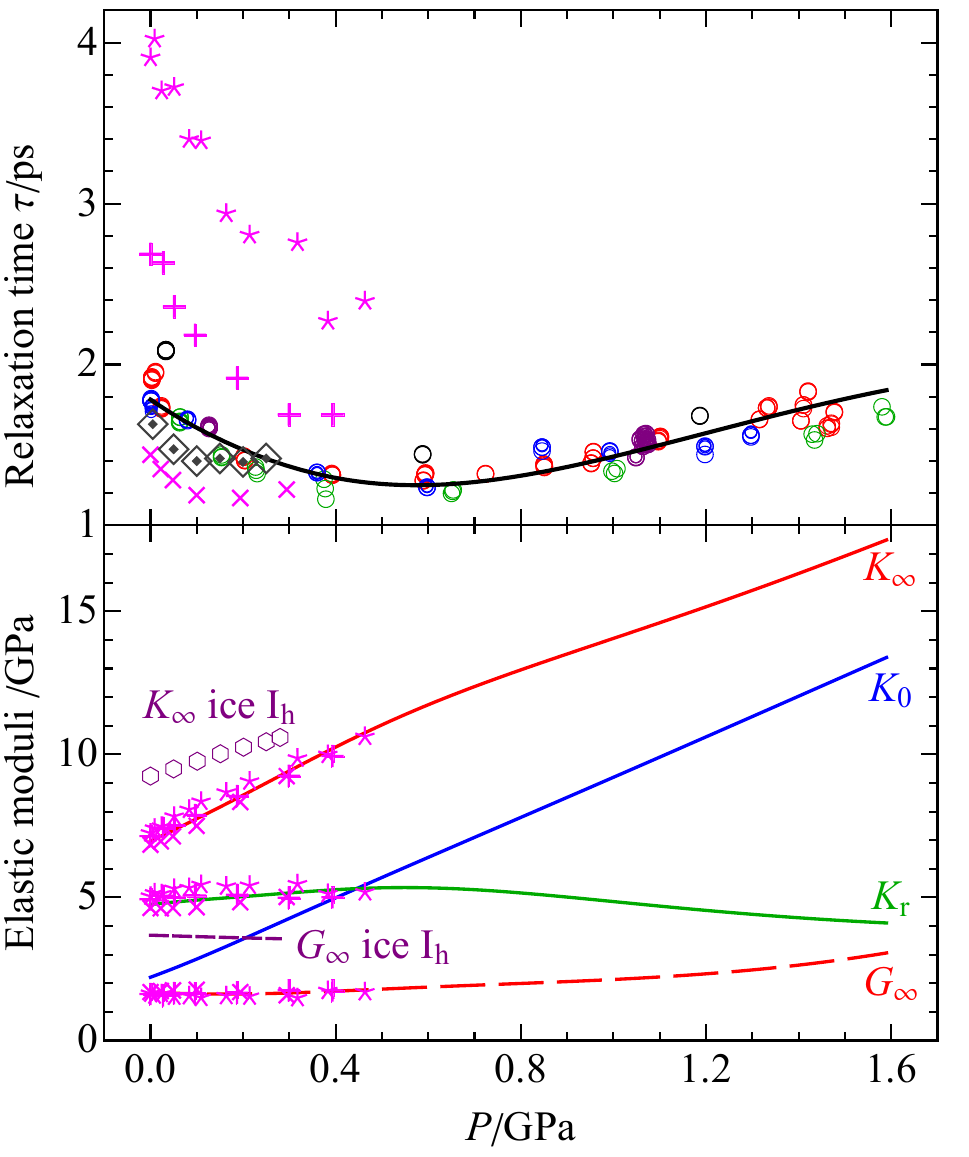}
\caption{\textbf{Visco-elastic analysis.} Top: comparison of relaxation times as a function of pressure. Analysis of the present Brillouin data with Eqs.~\ref{eq:visco1}-\ref{eq:visco3} for individual measurements (open circles for runs A (blue), B (red), C (black), D (purple), and E (green)) and using the smooth fits to $\eta$ (Fig.~\ref{fig:eta}) and $\alpha/f^2$ (Fig.~\ref{fig:attenuation}) (solid curve); analysis with Eqs.~\ref{eq:visco1}-\ref{eq:visco3} of ultrasonic data~\cite{hawley_ultrasonicabsorption_1970} (magenta asterisks, pluses, and crosses at \num{273}, \num{283.3}, and \qty{303}{K}, respectively); and rotational relaxation time from NMR at \qty{299}{K}~\cite{lang_high_1981} (grey pointed diamonds). Bottom: elastic moduli. The solid curves show $K_0$ from the equation of state~\cite{theinternationalassociationforthepropertiesofwaterandsteam_revised_2018}, and $K_\mathrm{r}$, $K_\infty$, and $G_\infty$ calculated from Eqs.~\ref{eq:visco1}-\ref{eq:visco3} and smooth fits to $\eta$ (Fig.~\ref{fig:eta}) and $\alpha/f^2$ (Fig.~\ref{fig:attenuation}), or from the analysis of ultrasonic data~\cite{hawley_ultrasonicabsorption_1970} (magenta asterisks, pluses, and crosses at \num{273}, \num{283.3}, and \qty{303}{K}, respectively). Adiabatic moduli for hexagonal ice~\cite{gagnon_pressure_1988} are also shown: $K_\infty$ (purple hexagons) and $G_\infty$ (dashed curve).
\label{fig:tau}
}
\end{figure}
\begin{eqnarray}
\rho {w_\infty}^2 & = & K_\infty + \frac{4}{3} G_\infty \, ,\label{eq:visco1}\\
\frac{\alpha}{f^2} & = & \frac{2\pi^2}{w_0} \frac{K_\mathrm{r}+\frac{4}{3}G_\infty}{K_\infty} \,\tau \, ,\label{eq:visco2}\\
\eta_\mathrm{b} & = & \frac{K_0 K_\mathrm{r}}{K_\infty} \,\tau \,\label{eq:visco3}
\end{eqnarray}
where $K_\mathrm{r}=K_\infty - K_0$ (SM). Note that water also possesses other dynamical degrees of freedom: the librational mode, and the cage mode, corresponding to oscillations of a molecule between its neighbors. As these modes involve much higher frequencies than visco-elastic relaxation~\cite{bertolini_stress_1995}, a single relaxation step is enough for the present analysis. Using Eqs.~\ref{eq:visco1}-\ref{eq:visco3} and $w_\infty$ from Ref.~\cite{krisch_pressure_2002}, we converted our experimental data into $\tau$ and elastic moduli (Fig.~\ref{fig:tau}). This involves a short extrapolation of $w_\infty$ above \qty{1.35}{GPa} (Fig.~S13); we checked that the results are insensitive to the chosen extrapolation (Fig.~S14).

We find $\tau$ in the picosecond range (Fig.~\ref{fig:tau}, top). We also analyzed ultrasonic data at \qty{31}{MHz}~\cite{hawley_ultrasonicabsorption_1970} with the same procedure. The results at neighboring temperatures nicely bracket our values. We find that, in contrast to other fluids, $\tau$ initially decreases with increasing pressure, for temperatures from \num{273} to \qty{303}{K}. This trend was also reported in an inelastic UV scattering study~\cite{bencivenga_temperature_2009} up to \qty{0.35}{GPa}, with similar values for $\tau$, provided that $G_\infty$ is taken into account. Our data at \qty{295}{K} extend to much higher pressure than previous works, and reveal that $\tau$ reaches a minimum around \qty{0.5}{GPa} before recovering a normal behavior at higher pressure. Our analysis of ultrasonic data~\cite{hawley_ultrasonicabsorption_1970} suggests that minima could occur around a similar pressure at lower temperatures. A closer look at the factors involved in the calculation of $\tau$ (Eqs.~\ref{eq:visco1}-\ref{eq:visco3}) shows the following: the anomalous initial decrease of $\tau(P)$ stems from the very steep initial decrease of $\alpha/f^2$, which itself arises from the rather flat variation of $\eta$ and $\eta_\mathrm{b}$ while $w$ increases~(Eq.~\ref{eq:attenuation}). The pressure variation changes sign when $\eta$ and $\eta_\mathrm{b}$ increase fast enough; the increase of $w_0/w_\infty$ also contributes.

The analysis also provides the elastic moduli (Fig.~\ref{fig:tau}, bottom). Our results are in excellent agreement with those derived from the ultrasonic data taken at much lower frequency~\cite{hawley_ultrasonicabsorption_1970}, although they were collected at three different temperatures and therefore yielded different $\tau$ values. This strongly supports the visco-elastic analysis. As $\eta_\mathrm{b}/\eta=K_\mathrm{r}/G_\infty$ (SM), the pressure trend of the elastic coefficients explains the decrease of the viscosity ratio shown in Fig.~\ref{fig:bulk}. The infinite frequency moduli, which reflects the solid like behavior of the liquid at high frequency, can be compared to the elastic moduli of hexagonal ice~\cite{gagnon_pressure_1988}: $K_\infty$ and $G_\infty$ in liquid water have the same pressure trend as ice, and are lower by around 20 and 50\% respectively. This is reasonable, as the crystal presents a fully developed tetrahedral network.

\textit{Discussion---}To our knowledge, such a minimum in the pressure dependence of the structural relaxation time has not been reported before for water. We have found one simulation study of two models of fused silica~\cite{furukawa_negative_2021}, which is, like water, a tetrahedral network-forming liquid, and shares many of water's anomalies. For one of the models, the shear-stress relaxation time $\tau_\alpha$ was found to go through a minimum as a function of density at low temperature. Interestingly, the minimum was observed at a density where the Si–O coordination number at the first coordination shell is nearly 5, a structural change which Angell~\textit{et al.} suggested to drive the diffusivity anomaly in simulated fused silicates~\cite{angell_pressure_1982}.

Could a structural change be the physical origin of the observed minimum in $\tau$ in water? In 1948, Hall proposed a microscopic mechanism for sound absorption in water~\cite{hall_origin_1948}. Hall's model is based on the assumption of two interconverting local structural motifs: one highly tetrahedral, low-density state, and another more disordered, higher-density state, favored by compression. In this view, the structural relaxation time $\tau$ is a measure of the interconversion rates. The two-state description of water has since received substantial support from experiments and simulations~\cite{gallo_water_2016}. It provides an explanation for many anomalies of water, and has been used for a quantitative description of its experimental thermodynamic~\cite{holten_entropy-driven_2012,caupin_thermodynamics_2019,duska_water_2020} and dynamic~\cite{monterodehijes_viscosity_2018} properties in a broad temperature and pressure range. Specifically, in conditions where one of the two states dominates, the behavior of water is similar to normal liquids. In some temperature and pressure range, the fraction of each state undergoes a fast change, which causes the anomalies. For instance, the rotational relaxation time $\tau_\theta$, measured by NMR~\cite{lang_high_1981,arnold_pressure_2002}, first anomalously decreases along isotherms, reaches a minimum, and recovers a normal, increasing behavior at high pressure. We display in Fig.~\ref{fig:tau} $\tau_\theta$ at \qty{299}{K}, which lie very close to our data, suggesting that $\tau$ and $\tau_\theta$ are controlled by similar processes, involving the breaking of hydrogen bonds~\cite{laage_molecular_2006}. Indeed, at ambient pressure, the relaxation time (\qty{1.7}{ps}) is close to the timescale of the slow component observed in 2D infrared spectroscopy (\qty{1.8}{ps}), that simulations attribute to hydrogen bond breaking and formation (equilibration)~\cite{asbury_water_2004}. However, the minima in $\tau_\theta$~\cite{lang_high_1981,arnold_pressure_2002} occur at a substantially lower pressure than the minimum in $\tau$. This is also the case for other extrema in water's properties: density maxima along isobars, shear-viscosity minima and self-diffusion maxima $D$ along isotherms.

\begin{figure}[!t]
\centering
\includegraphics[width=.95\linewidth]{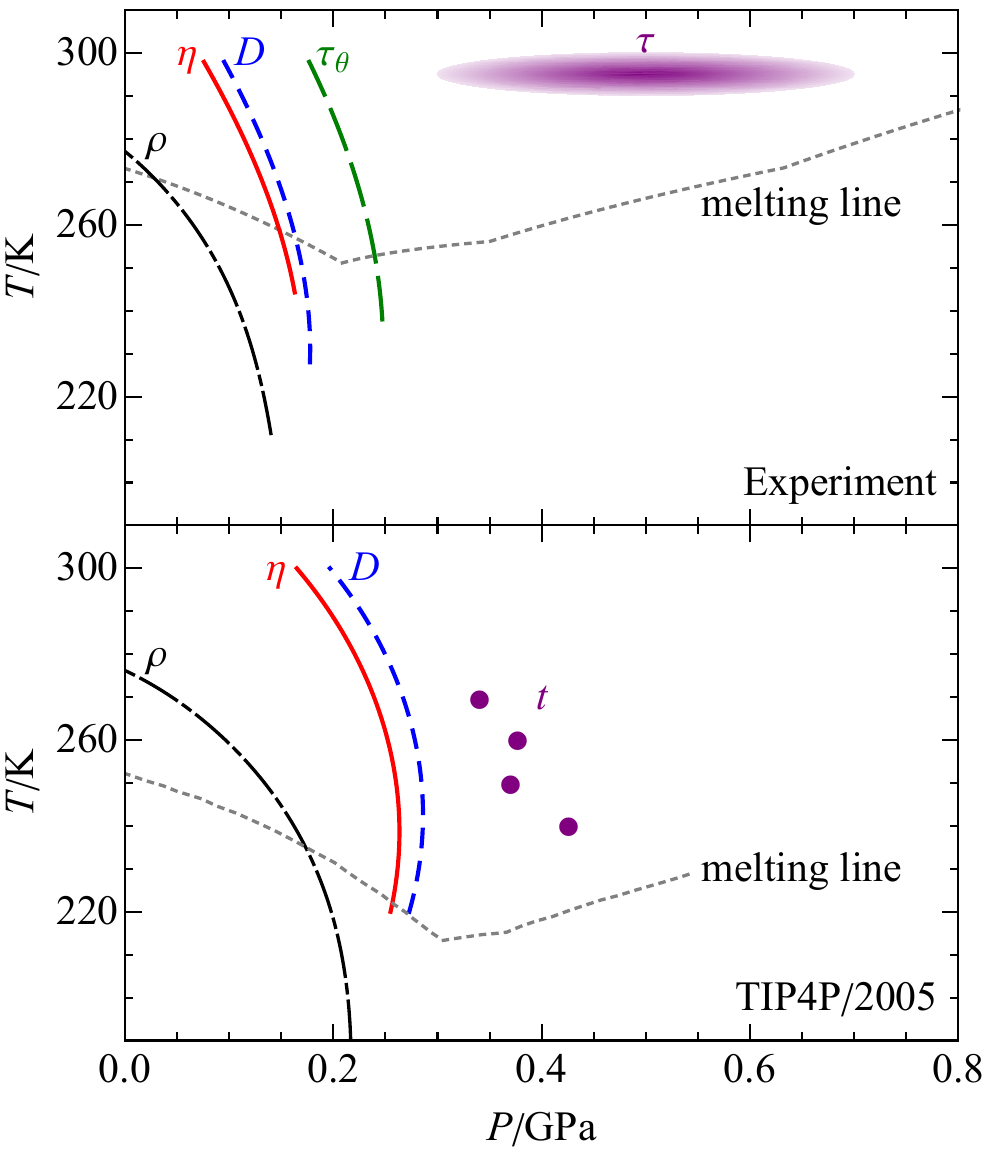}
\caption{\textbf{Lines of extrema in various properties of water.} Extrema along isobars for density $\rho$ (dashdotted black curve), and along isotherms for shear viscosity $\eta$ (full red curve), self-diffusion coefficient D (short-dashed blue curve), and rotational correlation time $\tau_\theta$ (long-dashed green curve). The extrema are calculated with two-state models for experiments~\cite{holten_entropy-driven_2012,singh_pressure_2017} (top) and simulations of TIP4P/2005 water~\cite{monterodehijes_viscosity_2018} (bottom). For real water, the purple ellipse in the top panel indicates the approximate location of the minimum in $\tau$ (see Fig.~\ref{fig:tau}). For simulations, the bottom panel displays minima of the translational order parameter $t$~\cite{agarwal_thermodynamic_2011} (purple discs). The gray dotted curves show the melting lines of the various ices for experiments~\cite{wagner_new_2011} and simulations~\cite{vega_what_2009}.
\label{fig:lines}
}
\end{figure}

The behavior of $\eta$, $D$, and $\tau_\theta$ was quantitatively explained by a dynamic two-state model~\cite{singh_pressure_2017}, which we use to display the lines of anomalies in Fig.~\ref{fig:lines}. The lines of extrema form a nested pattern. This feature was first studied for the SPC/E model of water by Errington and Debenedetti, who described it as a ``cascade of anomalies'': thermodynamic anomalies are embedded inside the region of dynamic anomalies, which are themselves embedded inside the region of structural anomalies. The latter has been identified in Ref.~\cite{errington_relationship_2001} using the translational order parameter $t$, which measures the tendency of pairs of molecules to adopt preferential separations in the liquid. The line of structural anomaly corresponds to minima in the pressure dependence of $t$ along isotherms. The cascade of anomalies nested inside the line of $t$ minima has been confirmed for other water models (mTIP3P, TIP4P, TIP4P/2005, and TIP5P~\cite{agarwal_thermodynamic_2011}), and also for Stillinger-Weber model of tetrahedral liquids tuned to mimic water, silicon, and germanium~\cite{dhabal_comparison_2016}. The case of TIP4P/2005 is shown for comparison in the lower panel of Fig.~\ref{fig:lines}: the similarity with real water is striking. The occurence of the structural anomaly could kinetically facilitate interconversion, causing the interconversion rate to reach its maximum, i.e. $\tau$ to reach its minimum. This suggests that the experimental anomaly in $\tau$, which occurs at higher pressure than any other thermodynamic or dynamic anomaly (Fig.~\ref{fig:lines}), is a good proxy for the structural anomaly in real water.

Concerning shear viscosity measurements, our results support previous data obtained with the rolling ball method~\cite{abramson_viscosity_2007} or with optical tweezers~\cite{bowman_optical_2013}, rather than a previous DDM attempt~\cite{frost_isotope_2020}. Nevertheless, our results show that DDM, when performed with an appropriate probe with a density close to the ambient fluid, is a reliable and efficient tool for viscosity measurements, that could be extended to other temperatures and other fluids.

\textit{Acknowledgments---}F.C. thanks M.A. Anisimov for helpful discussions. J.E., J.S., I.D., B.I., and F.C. were funded by Agence Nationale de la Recherche, Grant No. ANR-19-CE30-0035-01, and J.M.R. and C.V. by IHRC22/00002 and PID2022-140407NB-C21 from MINECO.

\clearpage
\centerline{\textsc{End Matter}}
\appendix
\section{Appendix A: Comparison with literature data for shear viscosity}

\begin{figure}[b]
\centering
\includegraphics[width=.95\linewidth]{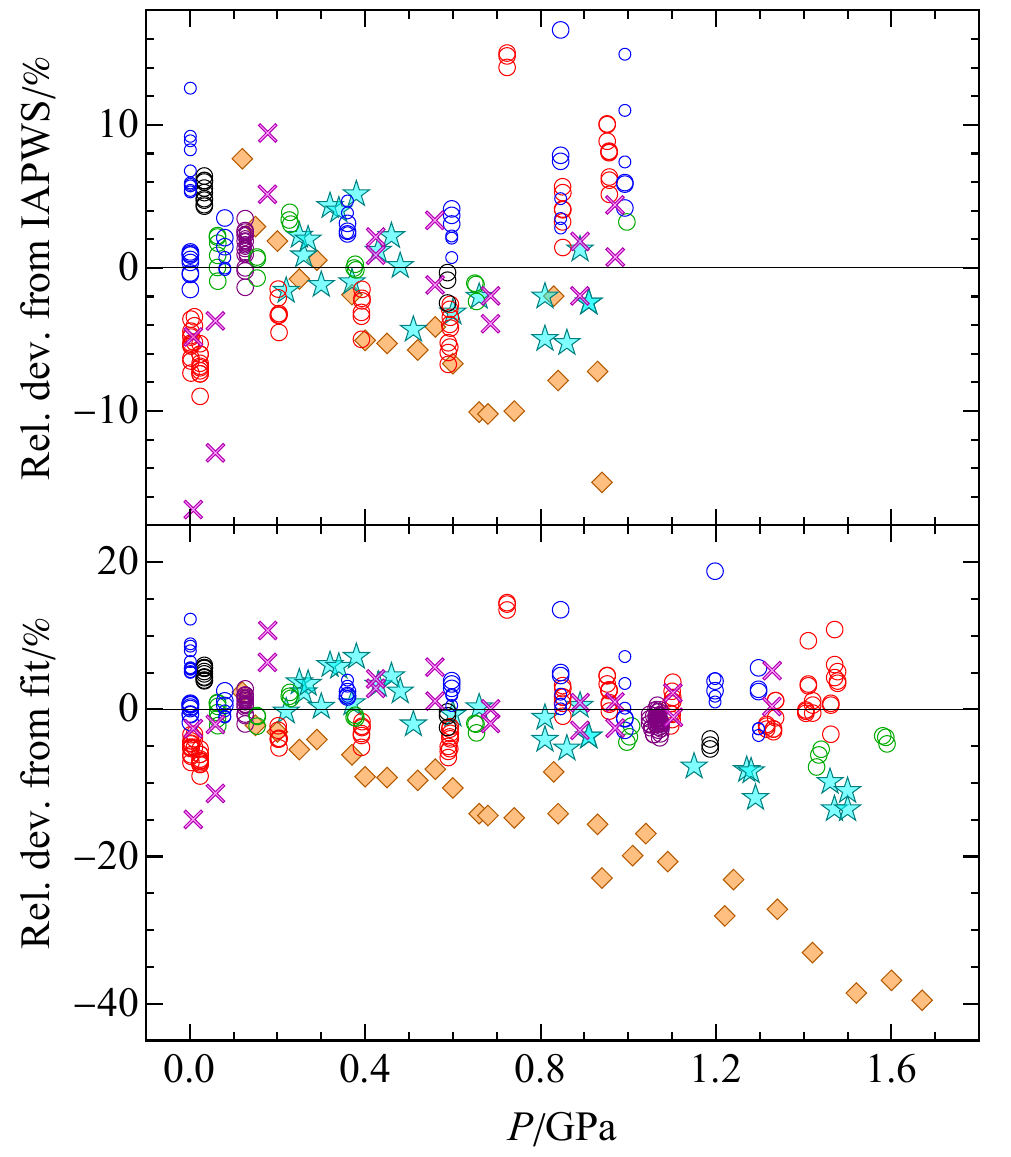}
\caption{\textbf{Comparison of shear viscosity data.} Top panel: percent deviation of the experimental data from IAPWS formulation; only data up to the limit of validity of the formulation (\qty{1}{\GPa}) are shown. Bottom panel: percent deviation of the experimental data from the fit to our data (see Fig.~\ref{fig:eta}). Our results are shown with open circles for runs A (blue), B (red), C (black), D (purple), and E (green). The large and small symbols stand for data obtained with 100x and 50x objectives, respectively. Also shown are the data obtained with the rolling ball method~\cite{abramson_viscosity_2007} (cyan stars), optical tweezers~\cite{bowman_optical_2013} (magenta crosses), and a previous DDM attempt~\cite{frost_isotope_2020} (orange diamonds).
\label{fig:etares}
}
\end{figure}

In Fig.~\ref{fig:etares}, we compare our data with various literature results. IAPWS formulation for viscosity~\cite{huber_new_2009} encompasses many references, with an estimated uncertainty of 1\% up to \qty{0.1}{\GPa}, and of 3\% from 0.1 to \qty{1}{\GPa}; it is not evaluated above \qty{1}{GPa}. Our data are consistent with this formulation up to \qty{0.7}{\GPa}, but become systematically higher at higher pressures, by up to around 7\% at \qty{1}{\GPa} (see Fig.~\ref{fig:eta}, middle panel). Abramson~\cite{abramson_viscosity_2007} measured shear viscosity in a DAC with the rolling ball method. His data are in good agreement with IAPWS up to \qty{1}{GPa}, but fall systematically below ours at higher pressure. We do not have a clear explanation for this moderate discrepancy. It would be worth repeating measurements with the rolling ball method, taking into account some precautions we suggest to avoid possible biases (SM). Bowman~\textit{et al.}~\cite{bowman_optical_2013} obtained shear viscosity from the positions fluctuations of a silica sphere trapped in a DAC by optical tweezers. Our data are fully consistent with this independent technique, which is also based on the principle of Brownian motion, but involves markedly different conditions. In our case, we use the light scattered by a large number of small, free spheres below the resolution of our microscope, whereas in Ref.~\cite{bowman_optical_2013} the image of a single, trapped \qty{5 }{\mu m} silica sphere is recorded. Importantly, silica is much less compressible than polystyrene, which, in contrast to our experiment, removes the need to account for a pressure-dependent radius (SM). The agreement between the two measurements thus gives confidence in our radius correction. Note that another measurement with a trapped polystyrene sphere was performed by the same group~\cite{saglimbeni_holographic_2016}, which gave data up to \qty{2}{GPa}. Unfortunately absorption of the IR trapping laser likely heated the sample above room temperature, and a temperature of \qty{307.15}{K} had to be assumed to match IAPWS formulation at low pressure.
\begin{figure}[b]
\centering
\includegraphics[width=.95\linewidth]{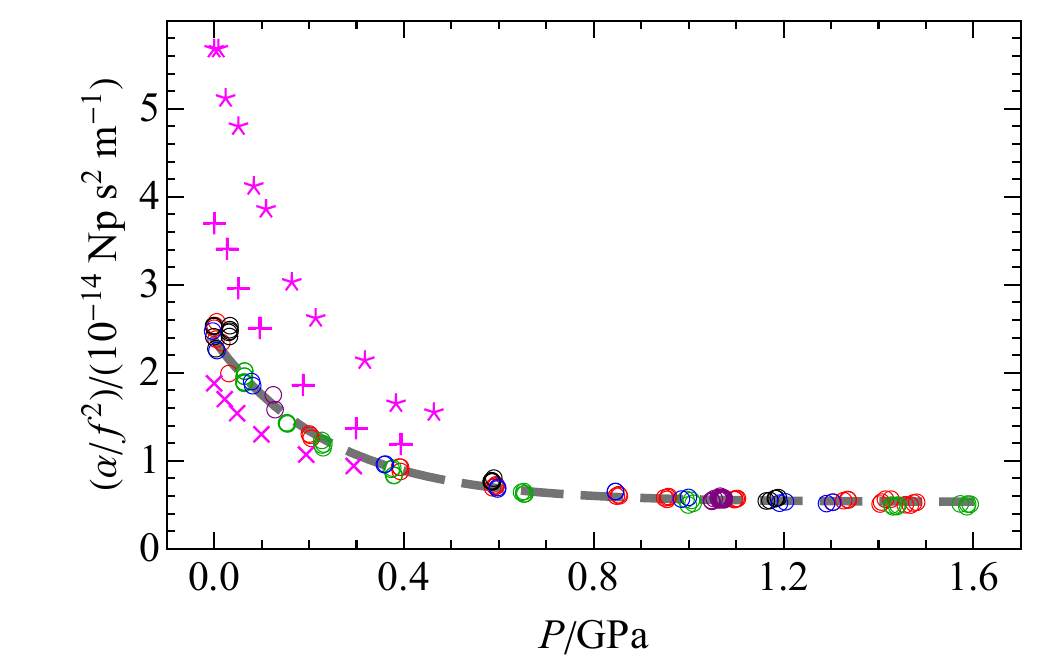}
\caption{\textbf{Sound attenuation $\alpha/f^2$ in water under pressure.} Our results are shown with open circles for runs A (blue), B (red), C (black), D (purple), and E (green). The thick, dashed, gray curve is a (vertically offset) exponential fit to all our data, whose parameters are given in SM. Ultrasonic data~\cite{hawley_ultrasonicabsorption_1970} at \num{273}, \num{283.3}, and \qty{303}{K} are shown with magenta asterisks, pluses, and crosses, respectively. 
\label{fig:attenuation}
}
\end{figure}
Finally, we consider another work, also based on DDM, by Frost and Glenzer~\cite{frost_isotope_2020}, who used \qty{400}{\nm} silica spheres. Their values are noticeably lower than all other works: they start to deviate downward from IAPWS at \qty{0.4}{GPa}, and the discrepancy with our data reaches -40\% at the highest pressure. We suspect that this is caused by rapid sedimentation of the spheres. Indeed, the density of silica is 2.6 times that of water, and the spheres are rather large. Their vertical motion may affect the DDM signal and accelerate the decorrelation between images, which would lead to an artificially low shear viscosity. We believe that our results and those of Bowman~\textit{et al.}~\cite{bowman_optical_2013} provide the most reliable values at high pressure. They show that the extrapolation of IAPWS formulation
\begin{figure}[t]
\centering
\includegraphics[width=.95\linewidth]{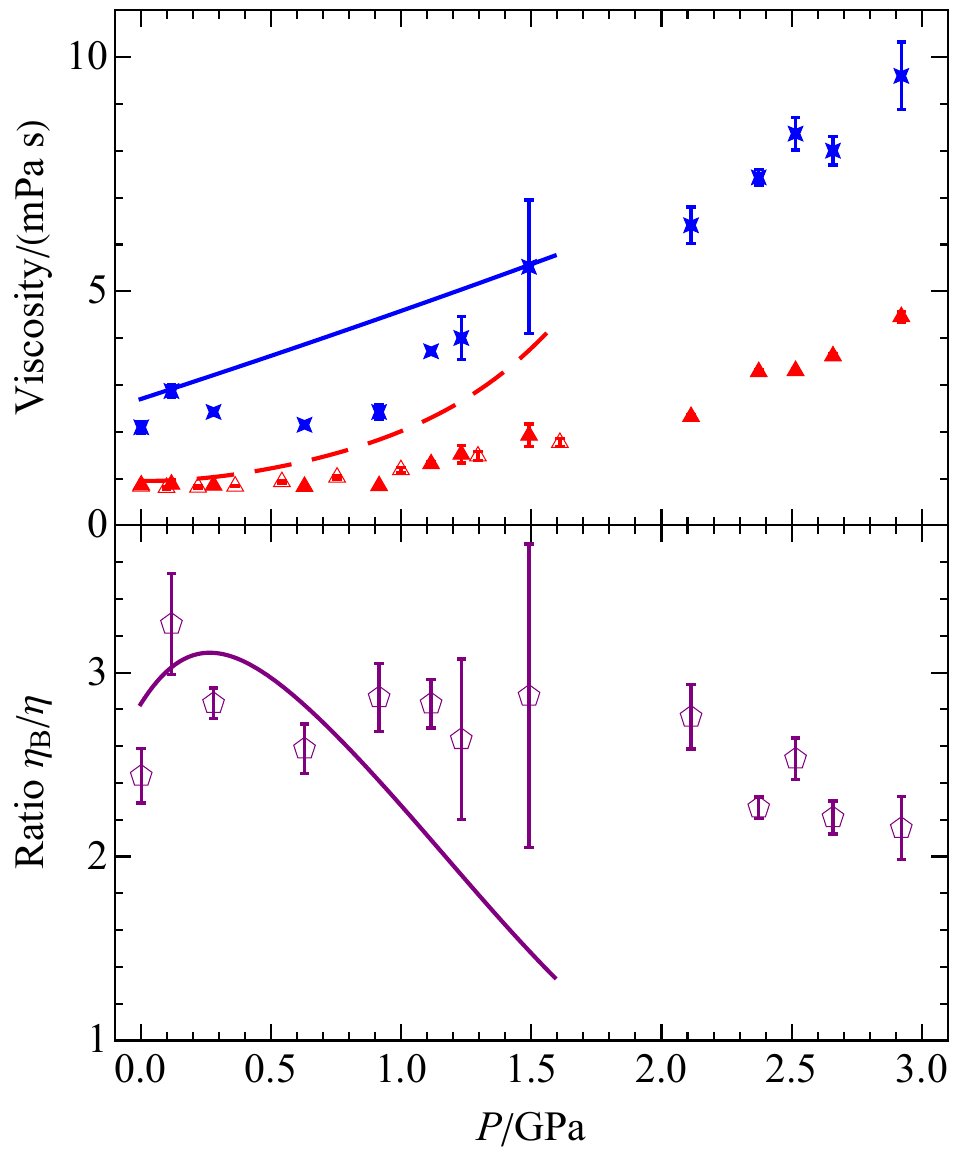}
\caption{\textbf{Simulation results for the TIP4P/2005 model of water at \qty{300}{K} as a function of pressure.} Top panel: shear (red triangles) and bulk (blue stars) viscosity. The empty symbols show previous simulation results~\cite{monterodehijes_viscosity_2018}. For comparison, the fits to the present experimental data are shown for shear (dashed red curve) and bulk (solid blue curve) viscosity. Bottom panel: ratio between bulk and shear viscosity for the simulations (pentagons). For comparison, the experimental curve (Fig.~\ref{fig:bulk}, bottom) is replicated here with a solid curve.
\label{fig:simul}
}
\end{figure}
beyond its range of validity (\qty{1}{GPa}) severely underestimates $\eta$, and therefore should not be used.

\section{Appendix B: Sound attenuation}

The amplitude of a sound wave propagating in a liquid decays exponentially at a rate $\alpha$ proportional to the squared frequency $f^2$. Figure~\ref{fig:attenuation} shows $\alpha/f^2$, obtained from the width of the Brillouin spectra (SM). The 5 runs are in good agreement with each other, and show an exponential decrease of $\alpha/f^2$ which plateaus above \qty{1}{GPa}. The standard deviation of the data around the exponential fit is 5.8\% . The ambient pressure value is consistent with literature data~\cite{holmes_temperature_2011}. We could find only two previous measurements of $\alpha/f^2$ in water under pressure: at \num{273.15} and \qty{303.15}{\K} up to \qty{0.2}{\GPa}~\cite{litovitz_effect_1955}, and at \num{273}, \num{283.3}, and \qty{303}{\K} up to \qty{0.4}{\GPa}~\cite{hawley_ultrasonicabsorption_1970}, both at low frequency. Figure~\ref{fig:attenuation} shows the latter work, whose data at the two temperatures closest to ours nicely bracket our values, which extend the covered pressure range by a factor 4.\\

\section{Appendix C: Simulations}

Whereas simulations have studied shear viscosity and its pressure dependence extensively, bulk viscosity has attracted much less attention, and available data are limited to pressures close to ambient~\cite{bertolini_stress_1995,jaeger_bulk_2018}. For this reason, we have performed extensive simulations of the shear and bulk viscosity along the \qty{300}{\K} isotherm, over a broad pressure range\cite{SM}. We have used the TIP4P/2005 model of water, which gives a shear viscosity at ambient conditions in excellent agreement with experiment. For shear viscosity, the present simulations agree with previous numerical results~\cite{monterodehijes_viscosity_2018}, and extend them to higher pressures (Fig.~\ref{fig:simul}). Although the agreement with experiment is excellent up to \qty{0.5}{\GPa}, the simulated values show too weak a pressure dependence. In order to reach shear viscosity values comparable to the highest measured (around \qty{4}{m\Pa\s}), we had to simulate up to \qty{3}{\GPa}. We also computed the bulk viscosity from simulations at the same state points (blue stars in Fig.~\ref{fig:simul}). The values agree well with the experimental ones, although simulations underestimate them around \qty{0.5}{\GPa}. The bottom panel of Fig.~\ref{fig:simul} shows the ratio between bulk and shear viscosity. The agreement with experiment is good up to \qty{1}{\GPa}, but at higher pressures the variation of the simulated ratio is milder than in the experiment. Eventually, above \qty{2}{\GPa}, the ratio starts decreasing with increasing pressure, as observed experimentally. Overall, the experimental trends are correctly captured by TIP4P/2005 simulations, although with a slower pressure variation.

\ifarXiv
    

\section*{Supplementary material}

Jan Eichler,[1, *] Johannes Stefanski,[1, *] José Martin Roca,[2] Isabelle
Daniel,[3] Bruno Issenmann,[1] Chantal Valeriani,[2] and Frédéric Caupin[1, †]

[1]*Institut Lumière Matière, Université de Lyon, Université Claude Bernard Lyon 1,*
*CNRS, Institut Universitaire de France, F-69622, Villeurbanne, France*
[2]*Departamento de estructura de la materia, física térmica y electrónica,*
*Facultad de Ciencias Físicas, Universidad Complutense de Madrid, 28040, Spain*
[3]*Laboratoire de Géologie de Lyon Terre, Planètes et Environnement,*
*Université Claude Bernard Lyon 1, CNRS, École Normale Supérieure de Lyon,*
*Université Jean Monnet Saint-Étienne F-69622, Villeurbanne, France*
(Dated: February 27, 2025)

## I. TEMPERATURE

The experiments were performed in an air-conditioned room. The temperature was measured next to the sample for each measurement with a calibrated platinum thermometer (Dostmann electronic, P795) with 0.015 K accuracy. The temperature of each individual measurement is given in the SI Datasets S1 to S12. The overall average temperature was 294.5 K, with a fluctuation within a day less than 0.5 K.

## II. BRILLOUIN SPECTROSCOPY, SOUND ATTENUATION, AND BULK VISCOSITY

### A. Brillouin spectroscopy

Brillouin light scattering (BLS) experiments were performed with the setup previously described [1]. In brief, 100 mW of a single longitudinal mode $\lambda = 532$ nm laser (Coherent Verdi V6) is focused on the sample through a microscope objective. Depending on the run, one of two Mitutoyo objectives was used: Plan Apo SL100x or 50x (both with N.A.=0.55). The backscattered light is routed to a Tandem Fabry-Pérot interferometer (JRS Scientific TFP-1), with 3 mm mirror spacing, corresponding to a free spectral range of 37.5 GHz. The entrance pinhole was either 150 or 200 μm in diameter, and the output pinhole was chosen accordingly: 200 or 300 μm, respectively.

For each setting, the instrument response function (IRF) was regularly determined by measuring the elastically scattered light from a piece of tape or teflon. It was found to be well represented by a Gaussian function $\propto \exp\left[-(\Delta f/c)^2\right]$, where $\Delta f$ is the frequency shift; values for the parameter $c$ were in the range 270 MHz to 330 MHz.

At each pressure, 3 to 4 Brillouin spectra were recorded, and each intensity $I(\Delta f)$ was fitted with a damped harmonic oscillator (DHO) function [2]:

$$I(\Delta f) = I_0 \frac{\Gamma_{\mathrm{B}}}{4} \left[ \frac{2 + \frac{\Delta f}{\sqrt{\Delta f_{\mathrm{B}}{}^2 - \Gamma_{\mathrm{B}}{}^2}}}{\left(\Delta f + \sqrt{\Delta f_{\mathrm{B}}{}^2 - \Gamma_{\mathrm{B}}{}^2}\right)^2 + \Gamma_{\mathrm{B}}{}^2} + \frac{2 - \frac{\Delta f}{\sqrt{\Delta f_{\mathrm{B}}{}^2 - \Gamma_{\mathrm{B}}{}^2}}}{\left(\Delta f - \sqrt{\Delta f_{\mathrm{B}}{}^2 - \Gamma_{\mathrm{B}}{}^2}\right)^2 + \Gamma_{\mathrm{B}}{}^2} \right], \tag{S1}$$

numerically convoluted with the Gaussian IRF. $I_0$ is an intensity factor, and $\Delta f_{\mathrm{B}}$ and $\Gamma_{\mathrm{B}}$ are the Brillouin frequency and half-width, respectively. Note that Eq. S1 rearranges into an equivalent, popular expression for the DHO:

$$I(\Delta f) = I_0 \frac{\Delta f_{\mathrm{B}}^2 \Gamma_{\mathrm{B}}}{(\Delta f^2 - \Delta f_{\mathrm{B}}^2)^2 + (2\Gamma_{\mathrm{B}}\Delta f)^2}, \tag{S2}$$

A frequency-independent baseline based on the dark count of the detector (0.39 s$^{-1}$) and the exposure time was taken into account. For better accuracy, the scattered intensity was binned in 36.7 MHz wide channels. The spectra reached

---

typically 100 counts at the peak. Using as uncertainty the square root of the counts for each point in the spectra, Eq. S1 gave excellent fits with a reduced $\chi^2$ always close to 1. An example is shown in Fig. S1.

To avoid systematic errors due to spectrometer calibration, and to compensate for the small daily drift of the spectrometer, we employed a relative measurement method, similar to the one introduced in Ref. [3]. Immediately before or after recording BLS in the diamond anvil cell (DAC), we recorded with the same setup 2 to 3 spectra from a test capillary (square $500\,\mu m$ cross-section) filled with water at ambient pressure. We thus obtained the ratio of the raw Brillouin frequency shift in the DAC to the average of the frequency shifts measured in the capillary, $r = \Delta f_{\mathrm{B,DAC}}^{\mathrm{raw}}/\Delta f_{\mathrm{B,CAP}}^{\mathrm{avg}}$, and used the known calculated value for water at ambient pressure $\Delta f_{\mathrm{B,CAP}}^{\mathrm{calc}}(T)$ to compute the absolute Brillouin frequency shift in the DAC: $\Delta f_{\mathrm{B,DAC}} = r\Delta f_{\mathrm{B,CAP}}^{\mathrm{calc}}(T)$. The raw Brillouin width in the DAC was corrected accordingly to give the absolute width: $\Gamma_{\mathrm{B,DAC}} = (\Delta f_{\mathrm{B,CAP}}^{\mathrm{calc}}(T)/\Delta f_{\mathrm{B,CAP}}^{\mathrm{raw}})\Gamma_{\mathrm{B,DAC}}^{\mathrm{raw}}$.

In a Brillouin experiment, the setup geometry defines a scattering angle $\theta$, which selects a wavevector $Q$:

$$Q = \frac{4\pi n}{\lambda}\sin\left(\frac{\theta}{2}\right)\,, \tag{S3}$$

where $n$ is the refractive index of the liquid at the laser wavelength $\lambda$. $\Delta f_{\mathrm{B}}$ is then related to the sound velocity in the liquid $w$ through:

$$\Delta f_{\mathrm{B}} = \frac{wQ}{2\pi} = \frac{2n}{\lambda}\sin\left(\frac{\theta}{2}\right)w\,, \tag{S4}$$

To compute the frequency shift for water at ambient pressure $\Delta f_{\mathrm{B,CAP}}^{\mathrm{calc}}(T)$, we set $\theta$ to $180°$ (backscattering geometry), and used formulations from the International Association for the Properties of Water and Steam (IAPWS) to compute $w$ [4] and $n$ [5]. Note that a precise knowledge of $\theta$ is not required, as the same scattering angle is used for the capillary and the DAC. We checked this assumption by measuring several spectra in the reference capillary and in water at ambient pressure through the top diamond of the DAC. The ratio between the average frequency shifts were found in the range 0.9993 to 1.0012, thus showing no significant difference, and no further correction was applied to the data.

### B. Refractive index of water under pressure

To convert the measured values of $\Delta f_{\mathrm{B}}$ and $\Gamma_{\mathrm{B}}$ into sound velocity and sound attenuation, we need the refractive index of the liquid. IAPWS provides a formulation for $n(\lambda,\rho)$ [5], but its range of validity is limited to densities less than $1060\,\mathrm{kg\,m^{-3}}$, which corresponds to less than $158\,\mathrm{MPa}$ at ambient temperature. To obtain values at higher pressure, we use the Gladstone-Dale (GD) formula:

$$n_{\mathrm{GD}}(\lambda,T,P) = 1 + K_{\mathrm{GD}}(\lambda,T)\,\rho(T,P)\,, \tag{S5}$$

where $\rho(T,P)$ is obtained from IAPWS-95 equation of state (EoS) [4], and the GD constant $K_{\mathrm{GD}}$ is calculated at ambient pressure $P_0 = 0.1\,\mathrm{MPa}$, using IAPWS formulations for $n[\lambda,\rho(T,P_0)]$ [5].

To check the validity of Eq. S5, we compare in Fig. S2 its prediction to actual data which cover the high pressure region, but were not included in IAPWS-95 formulation [6–10] (see Table S1 for details). The top panel shows the data used in the comparison, and the bottom panel displays the relative deviation from $n_{\mathrm{GD}}$ at the temperature and wavelength used in the corresponding experiment. Above $1\,\mathrm{GPa}$, we computed the density from the extrapolation of the IAPWS-95 EoS. We think this extrapolation is reliable in the range we need, as another EoS [11], based on supporting experimental data at higher pressures than IAPWS, differs from the extrapolation by only $+0.3\%$ in density at $300\,\mathrm{K}$ and $1.6\,\mathrm{GPa}$.

We find that the GD formula (Eq. S5) is compatible with the various data within their uncertainties (given in Table S1), except for Refs. [8] and [10]. We note however that these latter data may suffer from a systematic error, as the low pressure values of $n$ lie significantly below the values expected from the IAPWS formulation. Uncertainty on the pressure values may also contribute to some of the observed differences, as a typical $8\,\mathrm{MPa}$ pressure uncertainty converts into more than $0.1\%$ uncertainty on $n$. We conclude that Eq. S5 gives the refractive index with an accuracy better than $0.2\%$.

### C. Pressure from Brillouin spectroscopy

We have used the sound velocity $w$ from BLS to obtain the pressure $P$ in the liquid. To check the validity of the procedure, we carried out one run during which we simultaneously measured BLS and the R1 fluorescence line of

a ruby sphere within the DAC, which gives the pressure according to the equation by Mao *et al.* [12]. Figure S3 compares our results for $w(P)$ to literature data obtained with various methods involving lower sound frequencies: optical interferometry [13]; impulsive stimulated scattering [14]; Brillouin light scattering in platelet 80° scattering geometry [9] (see Table S2 for details). The pressure in these other works was also measured with ruby fluorescence [12], except for Ref. [13] where it was calculated from the geometry of and the pressure on a piston. The overall agreement is excellent. The data also agree with the sound velocity calculated with the IAPWS-95 formulation for the EoS of water [4], even when extrapolated beyond 1 GPa, its stated limit of validity. For a better comparison, the bottom panel of Fig. S3 displays the relative residuals between the experiments and the IAPWS-95 formulation. The average and root mean square of the relative residuals for our data are 0.51% and 0.91%, respectively. The average value of the relative error on sound velocity corresponds to a systematic error on pressure in the range 4 MPa to 6.5 MPa, below the typical 10 MPa uncertainty in pressure from ruby fluorescence.

Having thus shown the consistency of our BLS values of $w$ with thermodynamic values from the EoS, we can now invert Eq. S4 to turn Brillouin spectroscopy into a pressure sensor for experiments in pure water in a diamond anvil cell. We combine Eqs. S4 and S5 with the IAPWS-95 formulation [4] for $\rho(T, P)$, and solve the resulting equation:

$$\Delta f_{\text{B,DAC}} = \frac{2}{\lambda}[1 + K_{\text{GD}}(\lambda, T)\,\rho(T, P)]w(T, P),\tag{S6}$$

to obtain the pressure $P$ from the measured Brillouin frequency shift $\Delta f_{\text{B,DAC}}$. Based on the results presented in the previous section, we estimate the $1-\sigma$ absolute uncertainty on pressure to 10 MPa. Brillouin spectra were recorded before and after each DDM measurement, and the pressure was calculated from the average of the corresponding Brillouin frequencies.

### D. Sound attenuation coefficient

In the hydrodynamic regime, the amplitude of a planar sound wave decays exponentially with distance $z$ as $\exp(-\alpha z)$. The sound attenuation coefficient $\alpha$ scales as the squared acoustic frequency:

$$\frac{\alpha}{f^2} = \frac{2\pi^2}{\rho w^3}\left[\frac{4}{3}\eta + \eta_{\text{b}} + \frac{\kappa}{C_P}\left(\frac{C_P}{C_V} - 1\right)\right]\tag{S7}$$

where $\eta$ and $\eta_{\text{b}}$ are the shear and bulk viscosity, respectively, $C_P$ and $C_V$ the isobaric and isochoric heat capacity, respectively, and $\kappa$ is the thermal conductivity.

### E. Thermal contribution to sound attenuation

Because of the vicinity of the density maximum where $\gamma = C_P/C_V = 1$, the contribution of the last term between square brackets in Eq. S7 is usually neglected near room temperature. However, as we reach pressures above which the density becomes monotonic along isobars, we need to estimate the magnitude of this term. To this end, we use IAPWS formulations for shear viscosity [15] and thermal conductivity [16]. At ambient temperature, they are valid up to 1 GPa, where their uncertainties reach 3 and 5 %, respectively. Figure S4 displays a contour plot of the ratio between the last and first terms between square brackets in Eq. S7. We see that, at ambient temperature, the contribution from the thermal term does not exceed 1% of the one from the shear viscosity at any pressure. In addition, our results show that the contribution from the bulk viscosity to Eq. S7 is even larger than the one from the shear viscosity. Therefore, we can safely neglect the thermal term in Eq. S7.

### F. Bulk viscosity

Once we have measured $\eta$ with DDM, and $\Delta f_{\text{B}}$ and $\Gamma_{\text{B}}$ with BLS at the same conditions, we can deduce $\eta_{\text{B}}$. We start with the relation:

$$\alpha = 2\pi\frac{\Gamma_{\text{B}}}{w}.\tag{S8}$$

Then we use the pressure determined from BLS, and the corresponding $\rho$ and $n$ from IAPWS-95 EoS and from Eq. S5, combined with Eqs. S4 and S7 without the thermal term, to obtain:

$$\eta_{\text{b}} = \frac{\rho\lambda^2\Gamma_{\text{B}}}{4\pi n^2} - \frac{4}{3}\eta.\tag{S9}$$

### III. SHEAR VISCOSITY

#### A. DDM experiments

The experimental parameters for the present DDM measurements are similar to those used in our previous works [17, 18]. The colloidal probes are polystyrene spheres (Thermo Scientific 3350A Nanosphere Size Standards) with a certified mean diameter determined by transmission electron microscopy and traceable to NIST of $345(7)$ nm, $k = 2$. The particle size distribution is highly monodisperse: the manufacturer gives for the diameter standard deviation 6.5 nm, i.e. a coefficient of variation CV=1.9%. We dilute the probes to a volume fraction of $10^{-4}$ in ultrapure water. This low concentration ensures a sufficient light scattering signal, while avoiding interactions between particles. We verified in our previous work that the measured diffusion coefficients were identical within uncertainty for volume fractions $2.5\,10^{-5}$, $3.3\,10^{-5}$, and $10^{-4}$. Moreover, the results obtained for viscosity of pure water at ambient pressure as a function of temperature agree within uncertainty with independent measurement methods [17]. Here, additionally, we checked that the Brillouin spectra of water in the suspension with $10^{-4}$ volume fraction were identical to those measured in ultrapure water, showing that the dilute suspension does not affect the Brillouin results. The density of the spheres is $1050\,\mathrm{kg\,m^{-3}}$ close to that of water, which results in a strong slowing down of settling. Experiments with one sample were performed within less than 24 h. We checked that the diffusion coefficient measured by DDM under the same thermodynamic conditions remained the same over this period.

For each DDM measurement, series of microscopic images of the center plane of the DAC through a Mitutoyo Plan Apo SL100x or 50x (both with N.A.=0.55) objective were recorded with a CCD camera (Prosilica GX1050, 1,024 × 1,024 pixels²; Allied Vision Technologies) able to reach up to 112 fps. The frame rate was progressively reduced with increasing pressure, as higher viscosities were measured. A typical sequence of 500 images was processed using a home-written code (MATLAB; Mathworks) based on Fourier analysis as explained in Ref. [19]. Thanks to the very narrow size distribution of the colloidal probes, at each wave vector $q$, an image decorrelation function involving a simple exponential with a single time $t$ is sufficient to fit the data. The diffusion coefficient $D$ for a given movie was obtained by least-squares fitting of $\ln t$ vs. $\ln q$ by the function $y = -\ln D - 2x$. At each pressure point, at least 3 DDM movies were taken, and the pressure was determined from Brillouin spectra measured just before and after the DDM acquisition (see Section II C).

#### B. Principle of the method

The measured diffusion coefficient $D$ of the colloidal probes is converted into the shear viscosity $\eta$ of the solvent thanks to the Stokes-Einstein relation:

$$\eta = \frac{k_\mathrm{B}T}{6\pi D R_\mathrm{h}} \ . \tag{S10}$$

For this, we need the hydrodynamic radius $R_\mathrm{h}$ of the probes. In our previous work on supercooled water at ambient pressure [17, 18], we could rely on two simplifying facts: (i) the temperature variation of $R_\mathrm{h}$ can be neglected due to the very low expansion coefficient of polystyrene; (ii) the Oseen correction due to hydrodynamic interactions with the container walls does not depend on viscosity but only on the aspect ratio $R_\mathrm{h}/d$ where $d$ is the wall separation. The latter correction thus amounts to a constant factor, that can be absorbed in an effective value for $R_\mathrm{h}$. Based on these assumptions, we could obtain accurate relative viscosity measurements by measuring $D$ at a reference temperature $T_0$, and writing:

$$\eta(T) = \eta(T_0)\frac{T}{T_0}\frac{D(T_0)}{D(T)} \tag{S11}$$

which eliminates $R_\mathrm{h}$ from the calculations.

In the present experiments inside the DAC, we have to face three issues: (i) polystyrene is compressible and $R_\mathrm{h}$ changes with pressure; (ii) we cannot easily set the DAC to a single reference pressure; (iii) the wall separation $d$ varies with pressure because the gasket is progressively squeezed. Here we describe how we circumvented these issues.

#### C. Change in radius due to pressure

The effect of pressure on the radius of $5\,\mu\mathrm{m}$ polystyrene spheres has been investigated by Saglimbeni *et al.* [20]. Using digital holographic microscopy in a DAC, they measured the radius $R$ as a function of pressure $P$ up to 2.12 GPa.

They provide the functional form they used to fit their results, but not the best fit parameters. Therefore, we digitized the data points shown in their Fig. 3(b) and fitted them with the following equation:

$$R(P) = R_* + \Delta R e^{-(P-P_0)/(3C)} \,, \tag{S12}$$

where $P_0$ is the ambient pressure. We obtained $R_* = 1.896 \,\mu\text{m}$, $\Delta R = 0.366 \,\mu\text{m}$, and $C = 0.644 \,\text{GPa}$. We convert Eq. S12 into an expression for the shrinkage factor $\beta(P) = R(P)/R(P_0)$:

$$\beta(P) = 1 + B \left[ e^{-(P-P_0)/(3C)} - 1 \right] \tag{S13}$$

with $B = \Delta R/(R_* + \Delta R) = 0.1619$. The result is shown on Fig. S5.

We can compare $\beta(P)$ to the previous work of Bridgman who measured the compression $(\Delta z/z)(P)$ of a cylindrical polystyrene block under uniaxial stress $\sigma_{zz} = P$ with lateral constraints [21]. The corresponding stress-strain relation is:

$$\epsilon_{zz} = \frac{\text{d}z}{z} = -(1-2\nu)\frac{1+\nu}{1-\nu}\frac{\sigma_{zz}}{E} \tag{S14}$$

where $E$ and $\nu$ are the Young modulus and the Poisson ratio, respectively. For hydrostatic compression, when pressure increases by $\text{d}P$, the relative radius change is $\text{d}R/R = -\text{d}P/(3K)$, where $K$ is the bulk modulus. Using $E = 3(1-2\nu)K$ and assuming that $\nu$ does not depend on pressure gives the shrinkage factor corresponding to hydrostatic compression:

$$\beta(P) = \left[ 1 + \frac{\Delta z}{z}(P) \right]^{\frac{1-\nu}{1+\nu}}. \tag{S15}$$

Using Bridgman's Table 1 [21] and $\nu = 0.35$ [22], we obtain the values displayed in Fig. S5. The agreement with Saglimbeni *et al.* [20] is excellent at low pressure and slightly deteriorates above 1 GPa. This might be due to the different polymer samples used. Elastic properties of several polystyrene polymers under pressure have been investigated [23]. In particular, the bulk moduli were found nearly identical for molecular weights ranging from $35\,000$ to $350\,000$, which implies that $\beta$ should be sample independent. In contrast, Young moduli can differ by several percents, so that using Bridgman's data may introduce systematic errors; still, up to the highest pressure we reach in our experiment, the departure between the two data sets in Fig. S5 remains below 1.6%. In the following, we will use the fit to the data of Saglimbeni *et al.* (Eq. S13), which were obtained in conditions similar to those of our experiments.

### D. Setting the value of the nominal hydrodynamic radius

As we cannot easily set a given low pressure in the DAC, we use measurements at various low pressures to set the *nominal* hydrodynamic radius of the colloidal probes, that is its value $R_\text{h}^0$ at ambient pressure. To this end, we combine the Stokes-Einstein relation, Eq. S13, the pressure $P$ deduced from Brillouin spectroscopy, and the diffusion coefficient $D$ measured with DDM, together with water's shear viscosity values calculated with the IAPWS formulation [15] under the same temperature and pressure conditions:

$$R_\text{h}^0 = \frac{1}{\beta(P)}\frac{k_\text{B}T}{6\pi\eta(T,P)D} \,, \tag{S16}$$

where $k_\text{B}$ is Boltzmann's constant. We limit this procedure to data at $P \leq 100 \,\text{MPa}$, for which the accuracy of the IAPWS formulation for viscosity is highest (1%). Moreover, at room temperature, as $\eta$ as a function of pressure goes through a minimum, its maximum relative variation in the range $0 \,\text{MPa}$ to $100 \,\text{MPa}$ is limited to 0.6%. The results are displayed in Fig. S6 and summarized in Table S3. There are systematic variations between runs, and even within a run when the microscope objective is changed. Nevertheless, all results remain close to each other. We therefore take a common, fixed value for $R_\text{h}^0$, equal to the overall average $175.6 \,\text{nm}$. The standard deviation around the average is $9.2 \,\text{nm}$, corresponding to a relative deviation of 5%. This is slightly higher than the typical uncertainty for a DDM measurement [17, 24], but includes here both systematic and random errors. The result $175.6(92) \,\text{nm}$ is in excellent agreement with the manufacturer's value: $172.50(175) \,\text{nm}$ (the uncertainty is given here as one standard deviation). This shows that the no-slip boundary condition, which gives the factor 6 in Eq. S16, is appropriate for the present probes, which are much bigger than typical slip lengths.

### E. Oseen correction

In our work on supercooled water, we used capillaries with a small inner thickness $d = 20\,\mu m$ to reduce the probability of ice nucleation. Hydrodynamic interactions between the colloidal probe and the capillary wall decrease the diffusion coefficient parallel to the wall, $D_\parallel$, compared to the free diffusion coefficient given by the Stokes-Einstein relation, $D_{\text{SE}}$. In our DDM experiment which averages over the full height $d$ of the sample [17], this leads to:

$$\frac{D_\parallel}{D_{\text{SE}}} = 1 - \frac{9}{8}\frac{R_h}{d - 2R_h}\ln\left(\frac{d - R_h}{R_h}\right).\tag{S17}$$

As this correction is independent on shear viscosity, it can be absorbed in the apparent hydrodynamic radius, provided that $d/R_h$ remains constant. This condition was realized in supercooled water at ambient pressure, as borosilicate glass and polystyrene have a negligible thermal expansion coefficient. Here, in the DAC, the sample chamber height is set by the gasket thickness, whose initial value is $d = 100\,\mu m$, but decreases with applied pressure; we observed a minimal value around $80\,\mu m$. $R_h$ also decreases with increasing pressure, see Fig. S5. The correction from the measured $D_\parallel$ to $D_{\text{SE}}$ calculated with Eq. S17 remains small, and, most importantly, varies negligibly with pressure: from $+1.27\%$ to $+1.42\%$ when $P$ varies from 0 to 1.6 GPa. To a good approximation, we can therefore assume a constant Oseen correction and absorb it in the value of $R_h^0$. If we were to correct the above value of $R_h^0$ by $-1.3\%$, we would obtain 173.2(91) nm, in even better agreement with the manufacturer's value, 172.50(175) nm. In the following, we neglect the Oseen correction.

### F. Final formula for viscosity

Based on the above analysis, we obtain the final formula to convert our measurements of $D$ with DDM into the shear viscosity $\eta$ of water at any pressure $P$:

$$\eta(P) = \frac{k_B T}{6\pi\beta(P)R_h^0 D(P)},\tag{S18}$$

with $R_h^0 = 175.6\,\text{nm}$.

## IV. FITS TO OUR EXPERIMENTAL DATA

For $\eta$ (Fig. 1, top panel), $\alpha/f^2$ (Fig. 2), and $\eta_b$ (Fig. 3, top panel), we performed a least squares fit to all the data from our five runs. In each case, we forced the value at 0 MPa to match existing literature data. We took $\eta(295.16\,\text{K}, 0\,\text{MPa}) = 0.9542\,\text{mPa s}$ from the IAPWS formulation [15]. For $\alpha/f^2$ and $\eta_b$, the values at 295.16 K and 0 MPa were set to $0.0237\,\text{Npm}^{-1}\text{MHz}^{-2}$ and $2.70\,\text{mPa s}$, respectively, based on interpolation of ambient pressure data at other temperatures from Ref. [25]. The formulas and parameters for the three fits are given in Table S4. For the ratio $\eta_b/\eta$ (Fig. 3, bottom panel), no fit was performed: the ratio of the fits to $\eta_b$ and $\eta$ was used instead.

## V. DISCUSSION OF ABRAMSON'S MEASUREMENTS

Here we investigate possible reasons for the difference between Abramson's viscosity values [26] and the present work.

Abramson used the rolling-ball method to measure viscosity inside a DAC. A detailed description can be found in his work on supercritical oxygen [27] and in a book chapter [28]. The principle is to measure the speed $v$ of a rolling sphere when the cell is tilted at an angle $\theta$ from the horizontal. The expected relation between $v$ and $\theta$ is:

$$v = \left[\frac{2\gamma R^2(\rho_S - \rho_F)g}{9\eta}\right]\sin\theta,\tag{S19}$$

where $R$ and $\rho_S$ are the sphere radius and density, $\rho_F$ and $\eta$ the fluid density and shear viscosity, $g$ the gravity, and $\gamma$ is a correction factor accounting for the finite gap in which the sphere is confined. However, Abramson found it necessary to include a vertical offset $C$ to analyze his data [27]:

$$v = \left[\frac{2\gamma R^2(\rho_S - \rho_F)g}{9\eta}\right]\sin\theta - C.\tag{S20}$$

He attributed the need for the offset $C$ to the existence of "a force through the center of mass of the sphere with component $-\Phi$ in the direction of travel, independent of tilt angle and speed". However, in Ref. [27], he acknowledged that "the origin and exact nature of the empirically necessary force $\Phi$ is not known," while in Ref. [28] the parameter $C$ was "presumed to derive from frictional forces".

The role of friction has been addressed in the case of a soft sphere rolling on an incline in air [29]. Elastic deformation creates a contact area, over which the normal reaction force is distributed inhomogeneously, resulting in a total normal reaction $\overrightarrow{N}$ offset by a distance $\delta$ from the center of the sphere. Combined with the weight $\overrightarrow{W}$ and the tangential friction force $\overrightarrow{F}$, this may result in enough rolling resistance to bring the sphere to a stop, even on an incline with a finite tilt angle $\theta$. We extend this model to the case of the rolling ball viscometer. The elastic deformation of small platinum spheres under their own weight is negligible, so that we rather ascribe the offset $\delta$ of the normal reaction $\overrightarrow{N}$ to the roughness of the sphere (see Fig. S7). Indeed, Abramson obtained the spheres by freezing droplets of molten platinum, which "appeared spherical over most of their surface with, however, a characteristic small fringe where the solidification terminated", and for which a roughness of $50\,\mathrm{nm}$ "is not unreasonable" [27]. The sphere being immersed in a viscous liquid, we include the viscous drag and torque, written $\overrightarrow{D} = -\alpha\eta R \overrightarrow{v}$ and $\Gamma = -\beta\eta R^3\Omega$, respectively. Here $\Omega = v/R$ is the angular velocity without slipping, and $\alpha$ and $\beta$ are numerical coefficients. Their values for a sphere in an infinite fluid are $\alpha = 6\pi$ and $\beta = 8\pi$ (Stokes law); they differ in the vicinity of walls but their exact values are not needed, as they will be absorbed in the correction coefficient $\gamma$. Writing the force and torque balance for steady rolling down, one obtains:

$$v = \left[ \frac{2\gamma R^2 (\rho_S - \rho_F)g}{9\eta} \right] \left( \sin\theta - \frac{\delta}{R}\cos\theta \right) , \tag{S21}$$

with $\gamma = 6\pi/(\alpha + \beta)$.

To see if this could affect the viscosity values, we reanalyzed data taken from Fig. 1 of Ref. [26] by fitting them with Eq. S21 instead of Eq. S20 (Fig. S8). The two fits are barely distinguishable, but the presence of the cosine in Eq. S21 makes the factor between square brackets smaller. The ratio between old and new values of this factor is displayed in Fig. S9. However, this does not directly affect the viscosity values, because Abramson's measurements are relative: he calculated the viscosity from the inverse of the slope times a "constant of proportionality [which] for each sphere is dependent on the diameter and was calculated from the known viscosities of water at 21 °C and pressures less than $0.5\,\mathrm{GPa}$". It turns out that, the relative change in slope due to switching from Eqs. S19 to S21 is small (1.9% at most), and, furthermore, nearly the same at low and high pressure for a given sphere. This means that using Eq. S19 instead of Eq. S21 would slightly change the constant of proportionality, but negligibly modify the viscosity values. We note however that we could perform this test only for the two pairs of points (Fig. S8) for which the rolling speeds are given for water in Ref. [26]. Therefore, we repeated the analysis in the case of nitrogen, for which Fig. 4.10 of Ref. [28] provides more data on rolling speeds (Fig. S10). Again, the two fits are barely distinguishable, and a similar conclusion is reached: Eq. S21 induces a slight change in the coefficient, but it is nearly the same at all pressures (see Fig. S9), leading to a small correction of viscosity values (a change of $+2\%$ at most from low to high pressure).

The results obtained for $\delta/R$ when fitting with Eq. S21 are displayed in Fig. S11; they range from 2 to 7%. For a given sphere, the ratio $\delta/R$ should not vary with pressure. This is the case for water. The 34 and 41 $\mu$m diameter spheres give $\delta \simeq 0.75$ and $0.5\,\mu$m, respectively. For nitrogen, the data are more scattered, but fall in three groups sharing the same value, suggesting that they correspond to three different spheres. Ref. [30] simply mentions that the spheres were of diameter 30 $\mu$m to 60 $\mu$m, so that $\delta$ is between 0.5 and 2 $\mu$m. The connection between $\delta$ and roughness was made by Bikerman [31]. He measured the tilt angle at which smooth spheres stopped rolling on a series of rough substrates. If the substrate presents "hills" of height $h$ separated by a typical distance $\delta$, they prevent rolling until a critical tilt angle $\theta_c$ is reached, with:

$$\frac{h}{R} = 1 - \cos\theta_c \quad \text{and} \quad \frac{\delta}{R} = \sin\theta_c . \tag{S22}$$

The relation is the same for a rough sphere on a smooth substrate, as the platinum sphere on the diamond window in the rolling ball viscometer. The tilt angle at which the fit with Eq. S21 crosses $v = 0$ (Figs S8 and S10) can be interpreted as $\theta_c$. $\theta_c$ is reached when the weight $\overrightarrow{W}$ passes through the point of contact C (see Fig. S7), and its torque vanishes. Hence $\sin\theta_c = \delta/R$. This also explains why the fits for the same sphere at various pressures cross $v = 0$ at the same point (Fig. S8). For the 34 and 41 $\mu$m diameter spheres used in water, $h \simeq 17$ and $5\,\mathrm{nm}$, occuring typically every $\delta \simeq 0.75$ and $0.5\,\mu$m, respectively. For nitrogen, $h$ is a few tens of nanometers, depending on the sphere radius. These values are reasonable.

We conclude that accounting for rotation resistance due to roughness gives a better understanding of the rolling ball viscometer in a DAC. Due to the small size of the spheres used, even a standard roughness amplitude plays a role

in the velocity-angle relation, and Eq. S21 should be preferred to Eq. S20 for data analysis. This provides a physical explanation for the hitherto empirical offset of the rolling speed curves. Nevertheless, the correction for viscosity does not exceed a couple of percent, which is comparable to the variation between runs (4 %), and the mesurement standard deviation (1 %) [27].

Another source of bias in the DAC is the narrow gap in which the sphere is confined. Abramson tackled this question by repeating measurements under the same conditions but changing the ratio $G = w/R$ (Fig. 4 of Ref. [27] and Fig. 4.12 of Ref. [28]), where $w$ is the distance from the top of the sphere to the upper diamond surface. The rolling speed starts from its asymptotic value at large $G$, decreases rather abruptly from $G \simeq 4$ to 0.6, and eventually plateaus at a value around 10% lower than the value at large $G$. Abramson concluded that "typical gap-to-diameter ratios in the diamond-anvil cell were between one and two and (...) the variations in rolling speed engendered by slight displacements of the upper diamond are not expected to be significant". Yet, $G = 1.4$ for a 41 $\mu$m sphere and a 100 $\mu$m gasket; squeezing the gasket to 80 $\mu$m would reduce $G$ to 0.95, which may modify the speed by several percents. However, the location of this interval in $G$ might also depend on the liquid viscosity, whereas the tests were only performed in a low viscosity liquid, using methanol at ambient conditions with a tilt angle of 30 degree, with spheres 70 $\mu$m to 100 $\mu$m in diameter. We estimate the corresponding Reynolds number for large $G$ to Re $= (\rho_F v R)/\eta \simeq 0.7$. We note that, for the drag on a sphere in an infinite fluid, the relative deviation from Stokes law is $3\text{Re}/16$ [32], which gives 13% at Re $= 0.7$. In the experiments on water, Re takes much lower values: when pressure increases, Re decreases from 0.01 to 0.001 and from 0.02 to 0.002 for 34 and 41 $\mu$m diameter spheres, respectively. The effect of $G$ might thus differ from the tests. These questions should be carefully considered in future work with the rolling ball viscometer in a DAC.

## VI. SIMULATION DETAILS

We have studied high pressure water using full-atomistic simulations as in Ref. [33]. In our work we have implemented the TIP4P/2005 water (rigid) model [34] in the LAMMPS Molecular Dynamics package [35], being one of the best force fields available to simulate water in a broad temperature and pressure range. To be coherent with the definition of TIP4P/2005, we have used a 0.85 nm cutoff. Molecules are held rigid using the SHAKE algorithm [36].

Throughout the work, we have studied the system in an NVT ensemble, with $N = 216$ molecules, keeping the temperature constant at $T = 300\,\text{K}$ with a Nose-Hoover thermostat. Densities ranged from 997 to 1430 kgm$^{-3}$. All state points have been simulated in the NVT ensemble for 50 000 steps with a time step of 1 fs, far beyond their characteristic time to ensure equilibration.

After the first equilibration, 10 ns NVE simulation at the energy corresponding to the equilibrium temperature is performed to compute shear $\eta$ and bulk $\eta_b$ viscosity applying the Green-Kubo relations for the autocorrelation of the stress tensor components [37, 38], given by

$$\eta = \frac{V}{6k_B T} \int_0^\infty \sum_\alpha \sum_{\beta \neq \alpha} \langle P_{\alpha\beta}(t)\, P_{\alpha\beta}(0) \rangle \tag{S23}$$

$$\eta_b = \frac{V}{9k_B T} \int_0^\infty \sum_\alpha \sum_{\beta \neq \alpha} \langle \delta P_{\alpha\alpha}(t)\, \delta P_{\beta\beta}(0) \rangle \tag{S24}$$

where $V$ is the volume of the system, $T$ is the equilibrium temperature set by the NVT simulation, $k_B$ is Boltzmann constant, $P_{ij}$ are the components of the pressure tensor in the system, $\delta P_{ij} \equiv \frac{1}{3}\text{Tr}(P_{ij}) - P$ is the fluctuation of the pressure tensor components, where $P = \frac{1}{3}\langle\text{Tr}(P_{ij})\rangle$ is the average pressure of the system and $\langle\cdot\rangle$ represent the ensemble average. It was shown by Tazi *et al.* [39] that $\eta$ obtained with Eq. S23 in a system with $N = 256$ molecules is equal to $\eta$ deduced from the size-dependent analysis of self-diffusion measured in systems with $N = 64$ to 2048.

## VII. VISCO-ELASTIC ANALYSIS

### A. Visco-elastic model

We use the viscoelastic model described in Ref. [40]. It introduces adiabatic elastic moduli for the liquid: $K_0 = \rho w_0^2$ and $K_\infty$ are the bulk moduli at zero and infinite frequency, respectively, and $G_\infty$ the shear modulus at infinite frequency. We assume a single relaxation process: pressure and shear stress relax at constant volume with the same

characteristic time $\tau_V$. This assumption is supported by experiments [41, 42] and simulations [43]). The frequency-dependent viscosities are then:

$$\eta(\omega) = \frac{G_\infty \tau_V}{1 + (\omega \tau_V)^2} \quad \text{and} \quad \eta_b(\omega) = \frac{K_r \tau_V}{1 + (\omega \tau_V)^2} \, . \tag{S25}$$

For a sound wave with angular frequency $\omega$, it yields the following sound velocity square:

$$w^2 = \frac{1}{\rho} \frac{K_0 + \left(K_\infty + \frac{4}{3}G_\infty\right)(\omega \tau_V)^2}{1 + (\omega \tau_V)^2} \, . \tag{S26}$$

The sound attenuation coefficient is:

$$\frac{\alpha}{f^2} = \frac{2\pi^2}{w} \frac{K_r + \frac{4}{3}G_\infty}{K_0 + \left(K_\infty + \frac{4}{3}G_\infty\right)(\omega \tau_V)^2} \tau_V \, . \tag{S27}$$

The sound velocity at infinite frequency is:

$$\rho {w_\infty}^2 = K_\infty + \frac{4}{3}G_\infty \, . \tag{S28}$$

At low frequency, Eqs. S26 and S27 become to leading order:

$$w \simeq w_0 \tag{S29}$$

$$\frac{\alpha}{f^2} \simeq \frac{2\pi^2}{w_0} \frac{K_r + \frac{4}{3}G_\infty}{K_0} \tau_V = \frac{2\pi^2}{w_0} \frac{K_r + \frac{4}{3}G_\infty}{K_\infty} \tau \, , \tag{S30}$$

where we have introduced $\tau = (K_\infty/K_0)\tau_V$, the relaxation time at constant pressure. We also have $\eta_b/\eta = K_r/G_\infty$.

The values of $\tau$, $K_\infty$, $K_r$, and $G_\infty$ displayed in Fig. 5 of the main text are derived from Eqs. S29 and S30. These equations are valid in the limit $(\omega \tau_V)^2 \ll 1$ and $(K_\infty + 4G_\infty/3)(\omega \tau_V)^2 \ll K_0$. The latter condition is the most stringent of the two. It corresponds to $(w_\infty/w_0)^2(\omega \tau_V)^2 \ll 1$, which is fulfilled in our experiments: based on values from Fig. 5, $(w_\infty/w_0)^2(\omega \tau_V)^2$ varies from 0.3 to 4% when $P$ varies from 0 to 1.6 GPa. Also based on values from Fig. 5, we can estimate the relative deviations for $w$ and $\alpha/f^2$ when Eqs. S29 and S30 are used instead of Eqs. S26 and S27: Fig. S12 shows that they remain small in the investigated pressure range.

### B. Values of $w_\infty(P)$

The pressure dependence of $w_\infty$ has been studied by Krisch *et al.* using inelastic x-ray scattering [44]. Figure S13 shows the data obtained near room temperature. The temperature dependence of $w_\infty$ being weak, data at 277 K and 297 K can be combined. In the main text, we model their pressure dependence with a linear function (solid line in Fig. S13), and used its extrapolation to analyze data from 1.4 GPa to 1.6 GPa. Figure S13 shows an alternate fit with a parabolic function (dashed curve). We repeated the analysis of the relaxation time $\tau$ with this alternate fit, and obtained a modified version of Fig. 5 of the main text, Fig. S14. The differences are very minor, amounting mainly to a slight increase of the $\tau$ values for pressure above 1.4 GPa, which makes the pressure minimum in $\tau$ even more apparent.

### VIII. SUPPLEMENTARY FILES

The files `runA_Brillouin.txt`, `runB_Brillouin.txt`, `runC_Brillouin.txt`, and `runD_Brillouin.txt` give the individual Brillouin measurements for each run. Each file contains the following columns: temperature $T$, $\Delta f_B$, $\Gamma_B$, $\chi^2$ of the DHO fit, pressure $P$, density $\rho$, sound velocity $w$, and sound attenuation coefficient $\alpha/f^2$.

The files `runA_50x_DDM.txt`, `runA_100x_DDM.txt`, `runB_100x_DDM.txt`, `runC_100x_DDM.txt`, `runD_50x_DDM.txt`, and `runD_100x_DDM.txt` give the individual DDM measurements for each run. Data taken with different objectives are given in separate files. Each file contains the following columns: temperature $T$, average pressure $P$, pressure drift $\Delta P$, diffusion coefficient $D$, shear viscosity $\eta$, bulk viscosity $\eta_b$, ratio $\eta_b/\eta$, relaxation time at constant pressure

$\tau$, bulk modulus $K_0$, relaxational modulus $K_r$, bulk modulus at infinite frequency $K_\infty$, and shear modulus at infinite frequency $G_\infty$.

---

TABLE S1. Literature data for the refractive index of liquid water at high pressure and room temperature.

| First author | Ref. | $\lambda$ (nm) | $T$ (K) | Pressure range (MPa) | $1 - \sigma$ uncertainty (%) | Data presentation |
|---|---|---|---|---|---|---|
| IAPWS | [5] | $470 - 670$ | $273.15 - 333.15$ | $0 - 150$ | 0.015 | Equation |
| Vedam | [6] | 546.1 | 298.15 | $54 - 1103$ | 0.1[a] | Table |
| Dewaele | [7] | 632.8 | 298 | $50 - 1200$ | 0.1[b] | Equation |
| Weiss | [8] | 532, 633 | 296.15 | $0 - 250$ | 0.03 | Equation |
| Sanchez-Valle | [9] | 514.5 | 293 | $30 - 1170$ | 1 | Table |
| Kim | [10] | 532 | room | $5.5 - 1295$ | 0.01 | Table[c] |

[a] Estimated from Ref. [45].
[b] Our estimate from the data scatter in Fig. 6 of Ref. [8].
[c] Private communication from Kim *et al.*.

TABLE S2. Literature data for the sound velocity of liquid water at high pressure and room temperature.

| First author | Ref. | Frequency (MHz) | $T$ (K) | Pressure range (MPa) | $1 - \sigma$ uncertainty (%) | Data presentation |
|---|---|---|---|---|---|---|
| Heydemann | [13] | $10 - 12$ | 295.15 | $0 - 1200$ | 0.9 | Table |
| Abramson | [14] | 1000 | 295.15 | $690 - 1280$ | 1 | Table |
| Sanchez-Valle | [9] | $4000 - 7100$ | 293 | $90 - 1170$ | 0.5 | Table |

TABLE S3. Hydrodynamic radius of the colloidal probes deduced from DDM measurements below $100\,\text{MPa}$ and from the shear viscosity from IAPWS formulation [15]. These values are without the Oseen correction.

| Run name | Objective | Number of points | Mean $R_\text{h}^0$/nm | Std. dev. $\delta R_\text{h}^0$/nm |
|---|---|---|---|---|
| A | 50x | 13 | 185.2 | 6.4 |
| A | 100x | 12 | 176.6 | 2.4 |
| B | 100x | 23 | 165.1 | 2.4 |
| C | 100x | 10 | 184.9 | 1.4 |
| E | 100x | 6 | 177.2 | 2.2 |
| All | | 64 | 175.6 | 9.2 |

TABLE S4. Parameters of the least squares fits to our data for $\eta$ (Fig. 1), $\alpha/f^2$ (Fig. 2), $\eta_\text{b}$ (Fig. 3), and to $w_\infty$ from Ref. [44] (Fig. S13). Here $P_\text{ref} = 1\,\text{GPa}$.

| Quantity | Functional form | $a_0$ | $a_1$ | $a_2$ | $a_3$ | $a_4$ |
|---|---|---|---|---|---|---|
| $\eta/(\text{mPa s})$ | $\sum_{i=0}^{4} a_i \left(\dfrac{P}{P_\text{ref}}\right)^i$ | $9.542 \times 10^{-1}$ | $-1.196 \times 10^{-1}$ | 1.693 | $-1.074$ | $5.579 \times 10^{-1}$ |
| $\dfrac{\alpha}{f^2}/(10^{-14}\,\text{Np s}^2\,\text{m}^{-1})$ | $(a_0 - a_1) \exp\left(-a_2 \dfrac{P}{P_\text{ref}}\right) + a_1$ | 2.374 | $5.325 \times 10^{-1}$ | 4.122 | — | — |
| $\eta_\text{b}/(\text{mPa s})$ | $\sum_{i=0}^{2} a_i \left(\dfrac{P}{P_\text{ref}}\right)^i$ | 2.698 | 1.809 | $7.402 \times 10^{-2}$ | — | — |
| $w_\infty/(\text{m s}^{-1})$ | $\sum_{i=0}^{1} a_i \left(\dfrac{P}{P_\text{ref}}\right)^i$ | $3.032 \times 10^3$ | $6.443 \times 10^2$ | — | — | — |
| $w_\infty/(\text{m s}^{-1})$ | $\sum_{i=0}^{2} a_i \left(\dfrac{P}{P_\text{ref}}\right)^i$ | $2.980 \times 10^3$ | $9.242 \times 10^2$ | $-2.155 \times 10^{-1}$ | — | — |

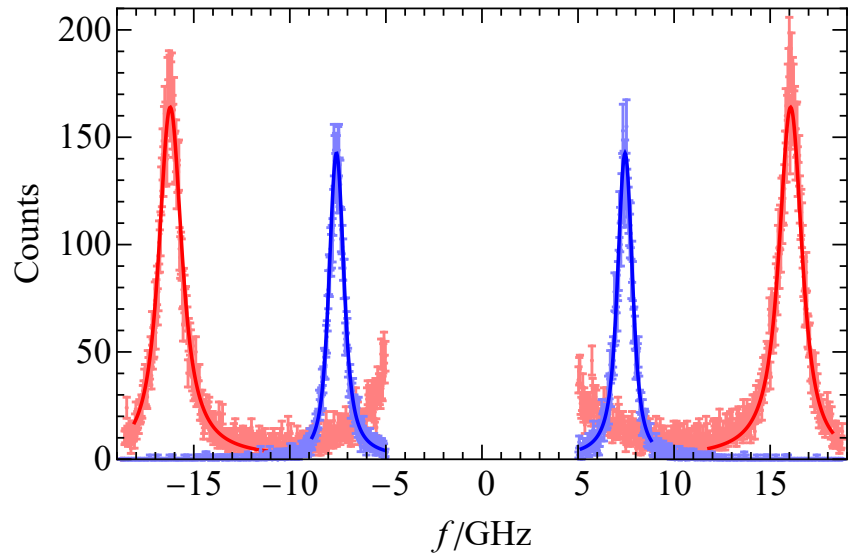

FIG. S1. Typical spectra: in water at $P = 1.29\,\mathrm{GPa}$ in the DAC for run A (red data at high frequency), and corresponding spectrum in water at atmospheric pressure in the reference capillary (blue data at low frequency). The curves show the fits with the DHO (Eq. S1) convoluted with the gaussian IRF.

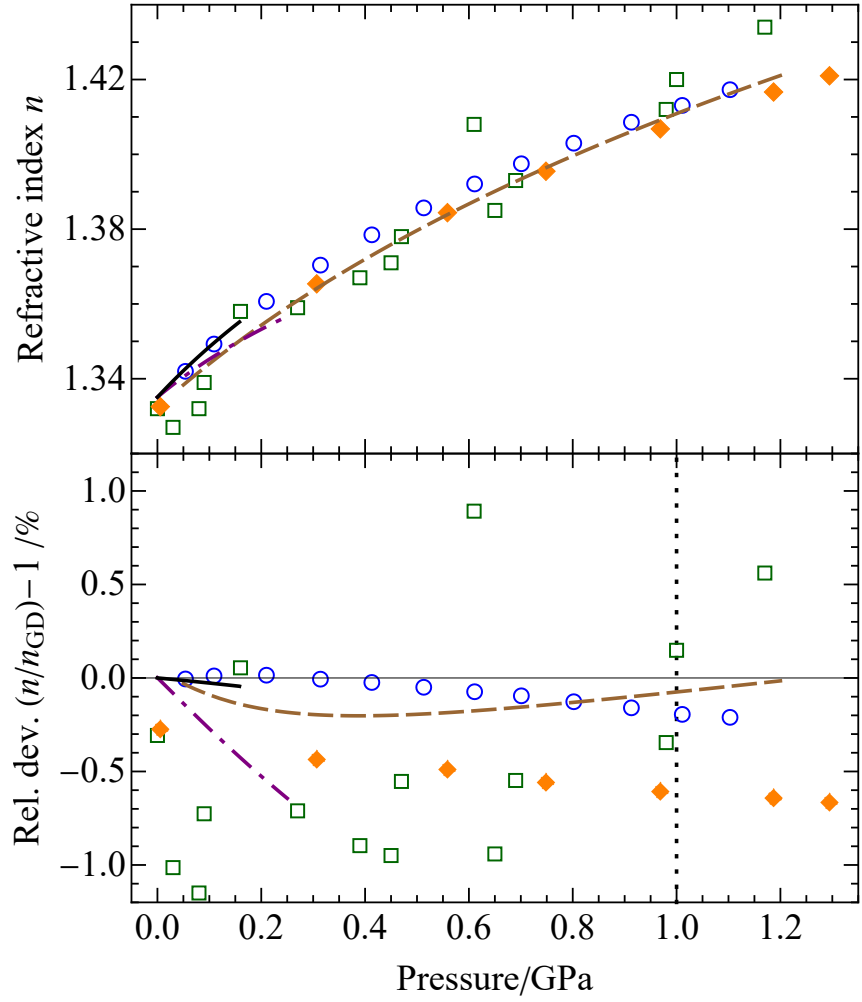

FIG. S2. Refractive index $n$ as a function of pressure $P$ (top) and percent deviation from the Gladstone-Dale expression, Eq. S5 (bottom). The symbols show the data from Refs. [6] (empty blue circles), [9] (empty green squares), [10] (filled orange diamonds), and the curves equations from Refs. [7] (dashed brown curve), [8] at 532 nm (dash-dotted purple curve), and the IAPWS formulation [5] at 532 nm and 296.15 K (solid black curve) (see Table S1 for details). The vertical, dotted black line shows the upper limit of validity of the IAPWS EoS [4] used to compute the density in Eq. S5.

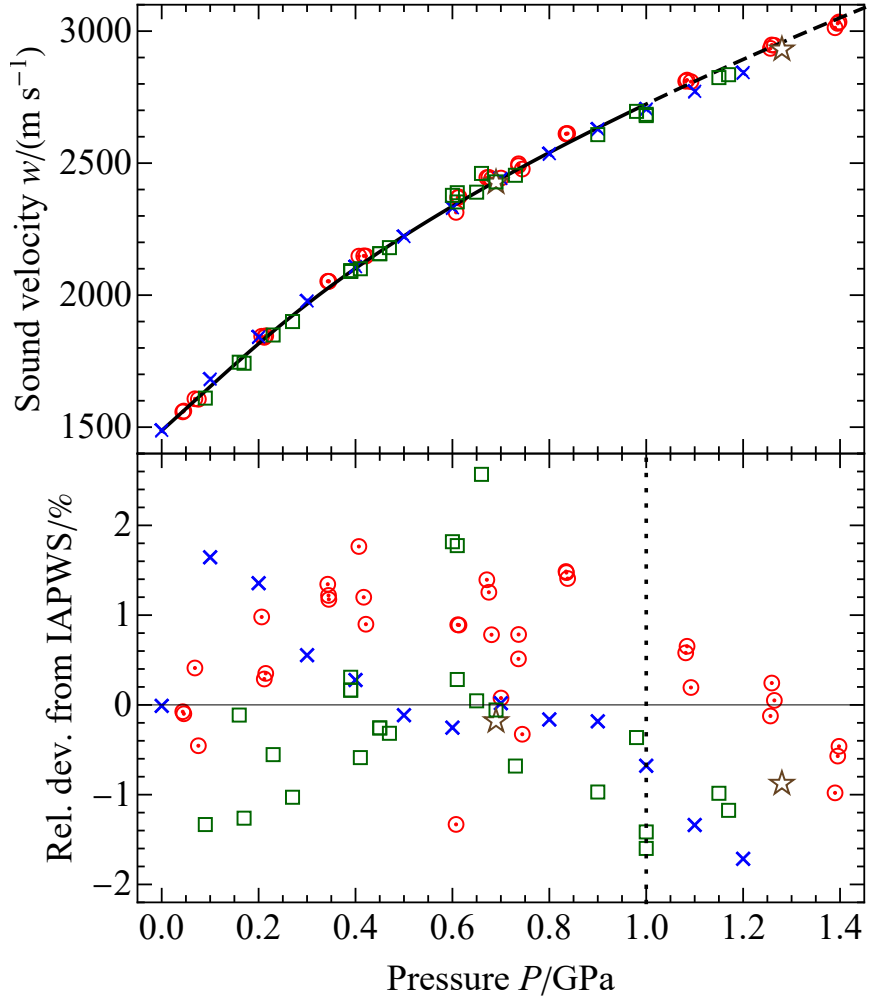

FIG. S3. Top: sound velocity as a function of pressure: this work, BLS at 180° and 294.65 K (pointed red circles); Refs. [13] (blue crosses), [14] (empty brown stars), [9] (empty green squares). The solid black curve is the sound velocity calculated with the IAPWS formulation [4] at 294.65 K; the dashed black curve shows its extrapolation beyond its stated limit of validity (1 GPa). Bottom: relative deviations between experimental values and the IAPWS formulation[4] (calculated at the corresponding temperature and pressure for each point) (same symbols at in top panel); the vertical dashed black line indicates the formulation's stated limit of validity.

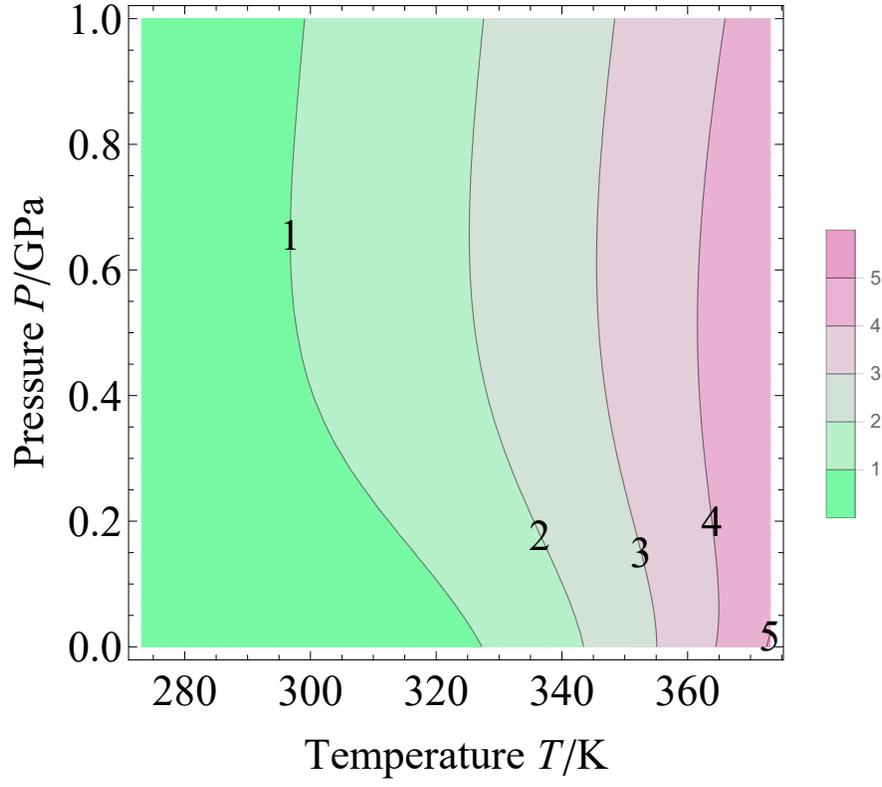

FIG. S4. Contour plot of the percent ratio between the thermal and shear viscosity contributions to sound attenuation (Eq. S7).

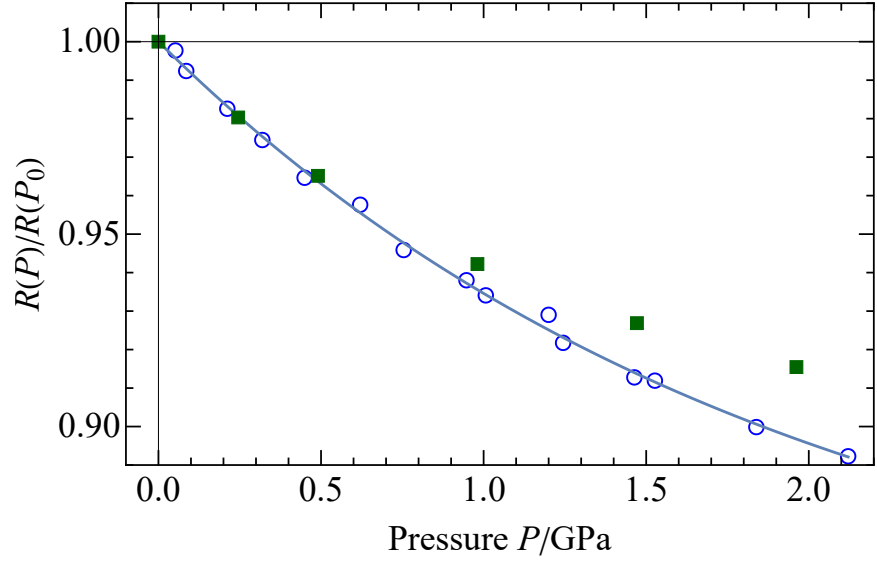

FIG. S5. Shrinkage factor $R(P)/R(P_0)$ as measured in Ref. [20] (open blue circles) and fitted with Eq. S13 (solid blue curve), or calculated from Bridgman's data [21] (filled green squares).

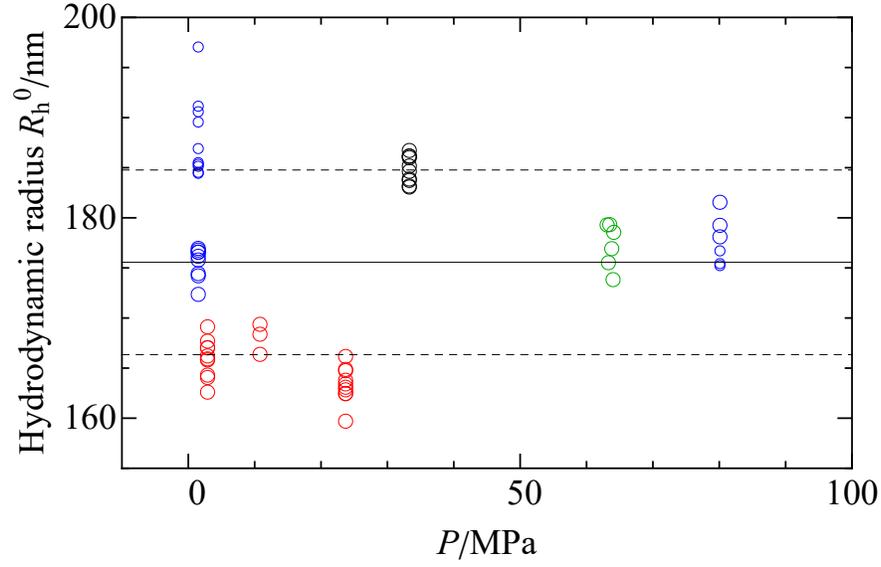

FIG. S6. Nominal hydrodynamic radius $R_h^0$ at ambient pressure as a function of the pressure at which $D$ was measured. Different runs are shown: A (blue), B (red), C (black), and E (green). All data were taken with a SL100x objective, except for the smaller symbols in run A which were taken with a 50x objective. The solid and dashed horizontal lines show the average value, and the values at plus or minus one standard deviation.

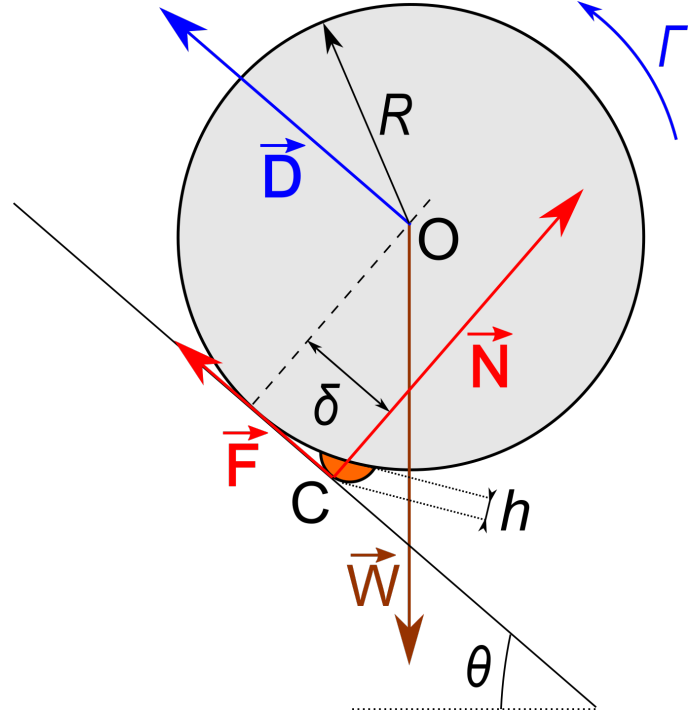

FIG. S7. Sketch of a sphere of radius $R$ rolling along an incline with tilt angle $\theta$. The roughness (shown here with only one orange "bump" of height $h$ for clarity) sets the point of contact C. The line of action of the normal reaction force $\vec{N}$ is shifted from the sphere center O by a distance $\delta$. The sphere is also submitted to its weight $\vec{W}$, the substrate friction $\vec{F}$, and the viscous drag $\vec{D}$ and torque $\Gamma$ from the fluid.

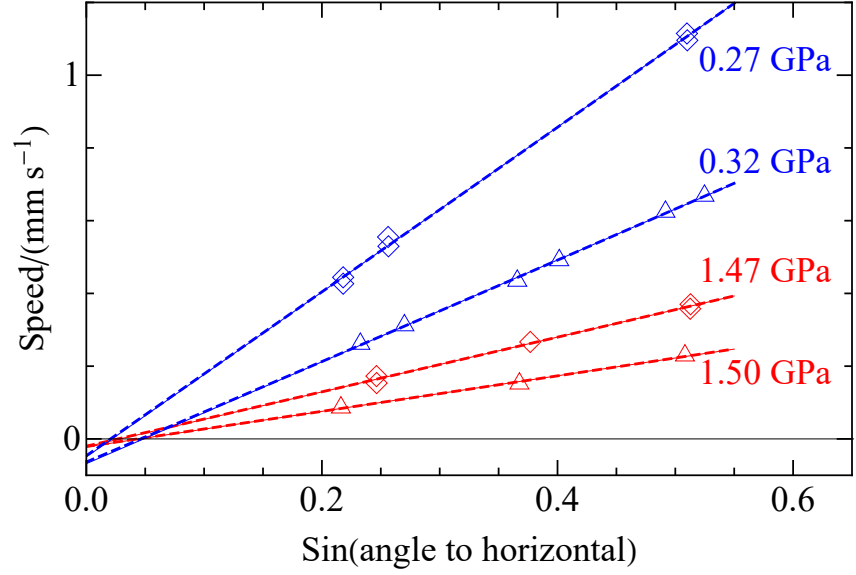

FIG. S8. Rolling speeds $v$ measured in water at 21 °C as a function of the sine of the angle to the horizontal. The data shown with triangles and diamonds were obtained with a sphere with 34 and 41 $\mu$m diameter, respectively. The labels give the pressure in GPa. The dashed and solid lines are fits with Eqs. S19 and S20, respectively.

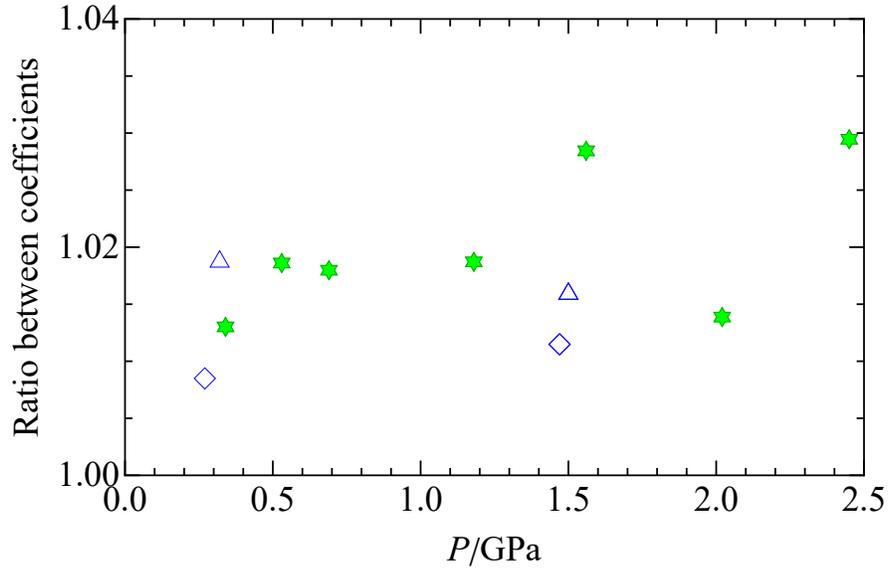

FIG. S9. Ratio between the factor between square brackets obtained with Eq. S20 and S21 as a function of pressure. The blue triangles and diamonds correspond to water and a sphere with 34 and 41 $\mu$m diameter, respectively. The green stars correspond to nitrogen and sphere diameters in the range 30 $\mu$m to 60 $\mu$m.

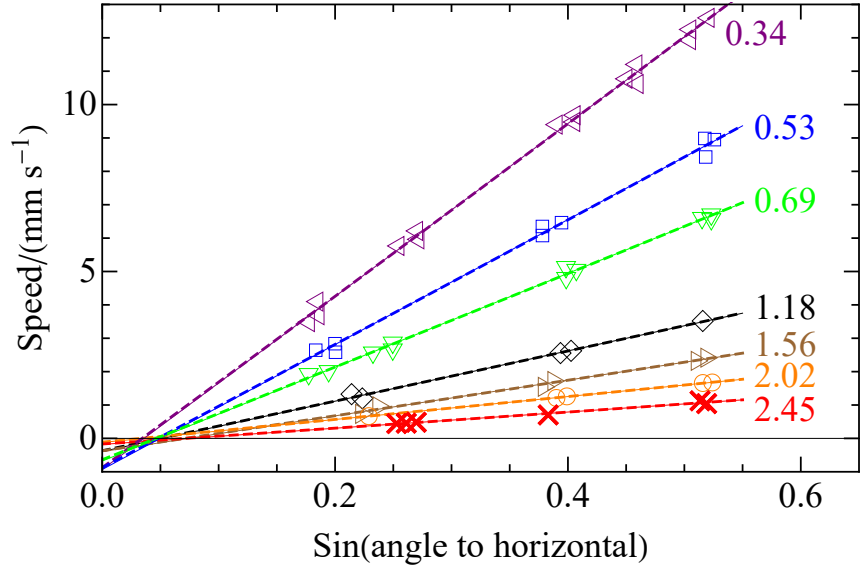

FIG. S10. Rolling speeds $v$ measured in nitrogen at 294 K as a function of the sine of the angle to the horizontal. The labels give the pressure in GPa. The dashed and solid lines are fits with Eqs. S19 and S20, respectively; the ratios between the respective slopes are shown in Fig S9.

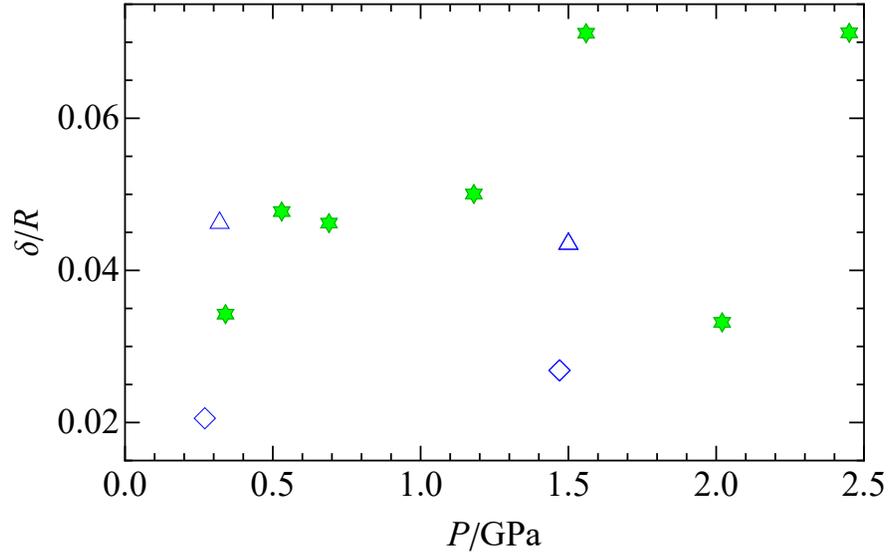

FIG. S11. Parameter $\delta/R$ obtained from the fit with Eq. S21 as a function of pressure. The blue triangles and diamonds correspond to water and a sphere with 34 and 41 μm diameter, respectively. The green stars correspond to nitrogen and sphere diameters in the range 30 μm to 60 μm.

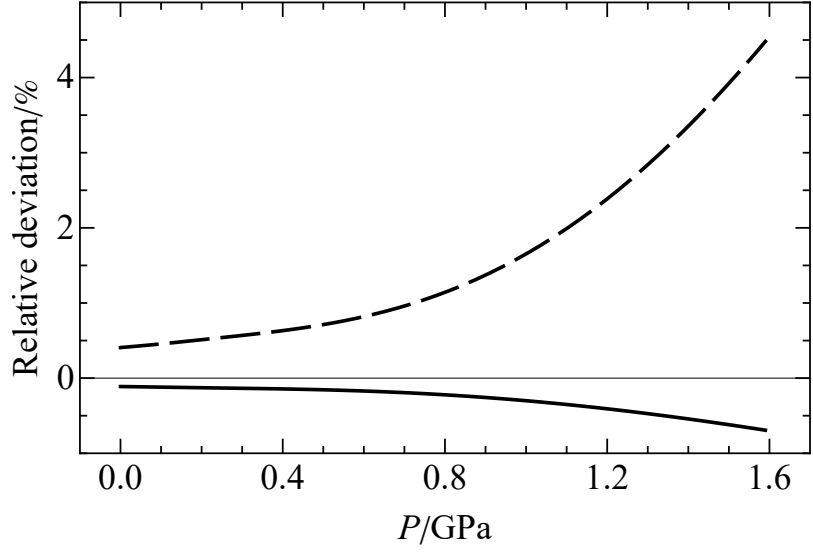

FIG. S12. Relative deviations for $w$ when Eq. S29 is used instead of Eq. S26 (solid curve), and for $\alpha/f^2$ when Eq. S30 is used instead of Eq. S27 (dashed curve).

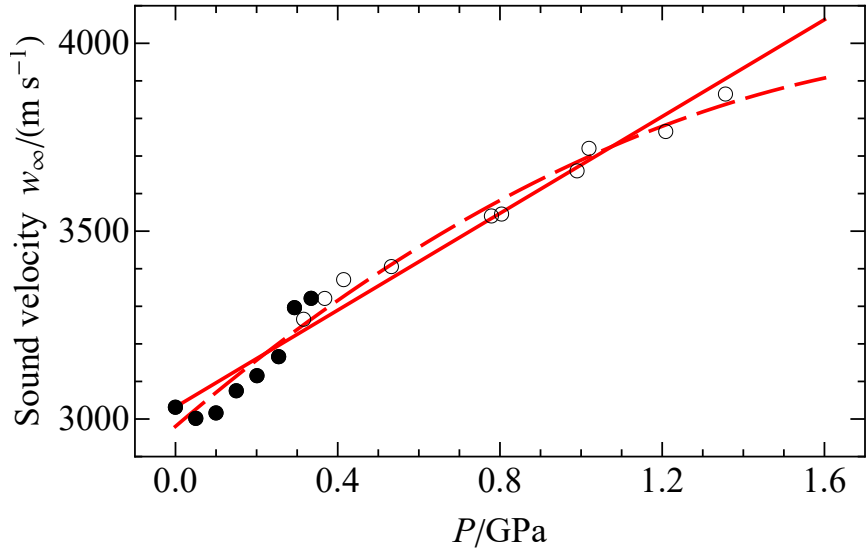

FIG. S13. High-frequency sound velocity $w_\infty$ as a function of pressure. Filled and empty circles show experimental values obtained by inelastic x-ray scattering [44] in a large volume cell at 277 K, and in a diamond anvil cell at 297 K, respectively. They are fitted with a straight line (solid curve), or a parabola (dashed curve).

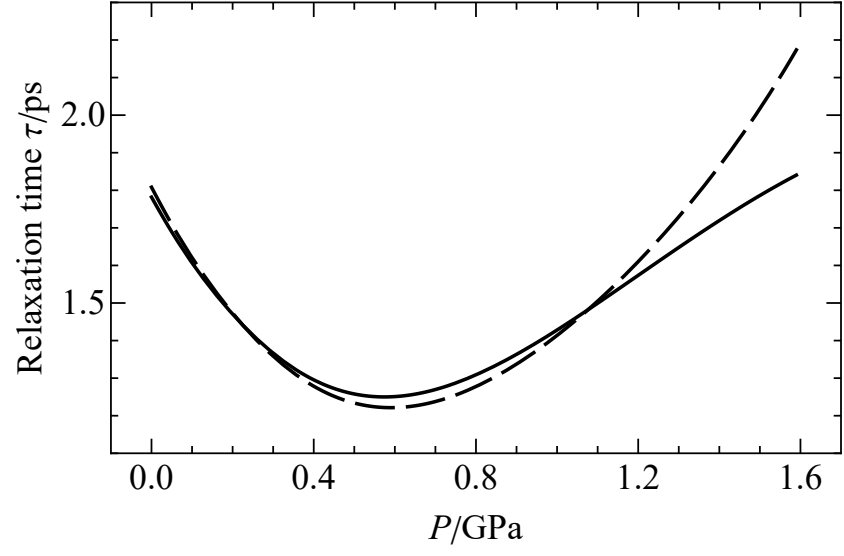

FIG. S14. Comparison of relaxation times as a function of pressure, obtained for two different fits of $w_\infty(P)$ (Fig. S13): linear (solid curve, as in Fig. 5 of the main text) and parabolic (dashed curve).


\fi

\end{document}